\documentclass[AMS,Times1COL]{WileyNJDv5} 

\usepackage{bm, mathtools, amsmath, amsfonts, amssymb, siunitx, subcaption}
\usepackage{amsthm}
\newtheorem*{rem}{Remark}
\usepackage[capitalise]{cleveref}
\usepackage{todonotes}
\newcommand{\Fmu}{\bm{F}_{\bm{\mu}}}
\newcommand{\detFmu}{\left|\det{\Fmu}\right|}
\newcommand{\Xmu}{\bm{x}^{\bm{\mu}}}
\newcommand{\Xp}{\bm{x}^{\text{p}}}
\newcommand{\Xmacro}{\macrobm{x}}
\newcommand{\Fmacro}{\macrobm{F}}
\newcommand{\Gmacro}{\widehat{\macrobm{\mathcal{G}}}}

\newcommand{\Omegamu}{\Omega^{\bm{\mu}}}
\newcommand{\Omegap}{\Omega^{\text{p}}}
\newcommand{\Phimu}{\bm{\Phi}_{\bm{\mu}}}
\DeclareMathOperator*{\argmax}{arg\,max}
\DeclareMathOperator*{\argmin}{arg\,min}
\newcommand{\dotthird}{\raisebox{-0.2em}{$\ \vdots \hspace{0.3em} $}}
\newcommand{\deriv}[2][\Xp]{\dfrac{\partial #2}{\partial #1}}
\newcommand{\derivmacro}[2][\Xmacro]{\bm{\nabla}_{#1} #2}
\newcommand{\inverse}[1]{#1^{-1}}
\newcommand{\norm}[1]{\left|\left|#1\right|\right|}
\newcommand{\trace}[1]{\operatorname{tr} #1}
\newcommand{\inverseT}[1]{#1^{-T}}
\newcommand{\macro}[1]{\overbar{#1}}
\newcommand{\macrobm}[1]{\macro{\bm{#1}}}
\renewcommand{\det}[1]{\operatorname{det} #1}

\newcommand{\diag}{\operatorname{diag}}
\newcommand{\integrate}[2][\text{p}]{\displaystyle\int_{\Omega^{#1}} #2 d\bm{x}^{#1}}
\newcommand{\integratemacro}[2][]{\displaystyle\int_{\macro{\Omega}^{#1}} #2 d\Xmacro^{#1}}
\newcommand{\meanintegrate}[2][\text{p}]{\frac{1}{\left|\Omega^{#1}\right|}\integrate[#1]{#2}}
\renewcommand{\wp}{{\bm{w}^{\text{p}}}}
\algrenewcommand\algorithmicrequire{\textbf{Input:}}
\algrenewcommand\algorithmicensure{\textbf{Output:}}
\algnewcommand\Or{\textbf{or}}
\newcommand{\expnumber}[2]{{#1}\times 10^{#2}}
\newcommand{\overbar}[1]{\mkern 1.5mu\overline{\mkern-1.5mu#1\mkern-1.5mu}\mkern 1.5mu}
\newcommand{\Wp}{\newtext{\bm{Y}^{\text{p}}}}
\newcommand{\Pp}{\bm{P}^{\text{p}}}
\newcommand{\Ap}{\bm{\mathfrak{A}}^{\text{p}}}

\definecolor{myred}{RGB}{222,45,38}
\definecolor{myblue}{RGB}{0,115,189}
\definecolor{mygreen}{RGB}{0,0,0}
\definecolor{gray}{RGB}{111,111,111}
\newcommand{\newtext}[1]{\textcolor{mygreen}{#1}}

\articletype{Article Type}

\received{Date Month Year}
\revised{Date Month Year}
\accepted{Date Month Year}
\journal{Journal}
\volume{00}
\copyyear{2023}
\startpage{1}

\begin{document}

\title{Reduced-order modeling for second-order computational homogenization with applications to geometrically parameterized elastomeric metamaterials}

\author[1,3]{T. Guo}

\author[2]{V. G. Kouznetsova}

\author[2]{M. G. D. Geers}

\author[1,3]{K. Veroy}

\author[2,3]{O. Roko\v{s}}

\address[1]{Centre for Analysis, Scientific Computing and Applications, Eindhoven University of Technology, 5612 AZ Eindhoven, The Netherlands}

\address[2]{Mechanics of Materials, Eindhoven University of Technology, 5612 AZ Eindhoven, The Netherlands}

\address[3]{Institute for Complex Molecular Systems, Eindhoven University of Technology, 5612 AZ Eindhoven, The Netherlands}

\corres{Theron Guo, Centre for Analysis, Scientific Computing and Applications,  Eindhoven University of Technology, 5612 AZ Eindhoven, The Netherlands. \email{t.guo@tue.nl}}

\fundingInfo{H2020 European Research Council, Grant/Award Number: 818473}

\abstract[Abstract]{The structural properties of mechanical metamaterials are typically studied with two-scale methods based on computational homogenization. Because such materials have a complex microstructure, enriched schemes such as second-order computational homogenization are required to fully capture their non-linear behavior, which arises from non-local interactions due to the buckling or patterning of the microstructure. In the two-scale formulation, the effective behavior of the microstructure is captured with a representative volume element (RVE), and a homogenized effective continuum is considered on the macroscale.

Although an effective continuum formulation is introduced, solving such two-scale models concurrently is still computationally demanding due to the many repeated solutions for each RVE at the microscale level. In this work, we propose a reduced-order model for the microscopic problem arising in second-order computational homogenization, using proper orthogonal decomposition and a novel hyperreduction method that is specifically tailored for this problem and inspired by the empirical cubature method. Two numerical examples are considered, in which the performance of the reduced-order model is carefully assessed by comparing its solutions with direct numerical simulations (entirely resolving the underlying microstructure) and the full second-order computational homogenization model. The reduced-order model is able to approximate the result of the full computational homogenization well, provided that the training data is representative for the problem at hand. Any remaining errors, when compared with the direct numerical simulation, can be attributed to the inherent approximation errors in the computational homogenization scheme. Regarding run times for one thread, speed-ups on the order of 100 are achieved with the reduced-order model as compared to direct numerical simulations.
}


\maketitle

\section{Introduction}\label{sec1}
With the recent advances in additive manufacturing, there has been a growing interest in designing and modelling metamaterials which exhibit emerging exotic properties that can be fine-tuned for specific applications. By a careful microstructural design, properties such as negative Poisson's ratio~\cite{Bertoldi2008MechanicallyStructures}, negative compressibility~\cite{Florijn2014ProgrammableMetamaterials}, or negative refractive index~\cite{Ramakrishna2005PhysicsMaterials} can be achieved. They can also act as filters that absorb certain bandwidths of frequencies~\cite{Krushynska2014TowardsMetamaterials}, or act as acoustic cloaks~\cite{Wang2022MechanicalDesign}. Additionally, such materials have been applied in impact mitigation~\cite{Tan2014Blast-waveMetamaterials} or biomedical applications~\cite{Greiner2012Micro-EngineeredStudies,Ning2018MechanicallyMesostructures}. A broad overview on their engineering applications can be found in~\cite{Surjadi2019MechanicalApplications}.

Studying properties of mechanical metamaterials through \newtext{direct numerical simulations (DNS)} is often challenging since complex microstructural geometries need to be resolved, requiring very fine meshes. In particular, in multi-query contexts such as the design of materials, numerous simulations are required and the computational costs become infeasible. To address this issue, multiscale methods based on computational homogenization (CH)~\cite{Feyel1999,Geers2010} are usually employed. These methods involve the separate modeling, discretization, and coupling of a microstructure defined on a representative volume element (RVE) and an effective homogenized macrostructure. The effective continuum does not resolve the complex microstructure on the macroscale but captures the underlying microscale physics. At every integration point of the macrostructure, the macroscopic kinematic quantities are used to specify the microscopic problem on the RVE which, after solution, returns effective quantities (e.g., stress and stiffness) back to the macroscopic solver. If scale separation can be assumed, i.e., the microstructural features are much smaller than the size of the macrostructure, the effective behavior of the microstructure can be adequately predicted using first-order CH~\cite{Kouznetsova2001,Miehe2002}. However, for metamaterials the microstructure can be of comparable size with the macrostructure and non-local effects (due to, e.g., buckling; see, e.g., ~\cite{Nguyen2014,Rokos2019a}) may emerge at the microscale, both violating the scale separation assumptions of the first-order scheme.

Enriched CH methods, such as second-order CH~\cite{Kouznetsova2004} or micromorphic CH~\cite{Forest2002Homogenization2}, extend the first-order formulation by introducing additional field variables and equations. For micromorphic CH, additional fields that describe the governing behavior of the underlying microstructure are introduced at the macroscale and communicated between both scales. To determine the evolution of these quantities, additional equations need to be included and a coupled system is solved at the macroscale. As an example, the average strain of the inclusions inside a composite RVE was introduced as an additional quantity in Biswas and Poh~\cite{Biswas2017}. In J\"{a}nicke et al.~\cite{Janicke2009Two-scaleScheme}, micro-rotations were considered as additional field variables for cellular materials. For buckling elastomeric metamaterials, prior knowledge on the buckling modes was embedded into the micromorphic framework presented in Roko\v{s} et al.~\cite{Rokos2019a}. For second-order CH, a strain gradient formulation is considered at the macroscale, i.e., the gradient of the strain (or deformation gradient) is required, giving rise to a length-scale associated with the size of the underlying RVE, thus making it possible to capture size and non-local effects. To ensure a proper scale transition of the kinematical quantities, additional constraints were derived in Kouznetsova et al.~\cite{Kouznetsova2004}. However, this model leads to artificial stress concentrations at the corners of the RVE for which subsequent formulations attempted to correct~\cite{Luscher2010AMaterials,Wu2023Second-orderMetamaterials,Yvonnet2020ComputationalBehavior}. In Luscher et al.~\cite{Luscher2010AMaterials}, additional constraints were derived from orthogonality conditions on the different components of the displacement field on the RVE, and in Wu et al.~\cite{Wu2023Second-orderMetamaterials} and Yvonnet et al.~\cite{Yvonnet2020ComputationalBehavior} body forces were included to account for additional effects.

In spite of the simplified formulation in terms of an effective non-local continuum, the multiscale problem based on CH is still computationally expensive, as the microscopic problem needs to be solved repeatedly. \newtext{Compared to the DNS, the two-scale model has much fewer degrees of freedom and elements on the macroscale. However, the evaluation of the constitutive model, i.e., solving the microscopic problem, is much more expensive. As a matter of fact, it might take longer to compute the multiscale model as compared to the DNS for small to intermediate scale ratios.} To overcome this problem, a reduced-order model (ROM) for the microscopic problem that is accurate and fast to evaluate is necessary. For first-order CH, numerous methods have been proposed in the literature. These methods can be split into two main classes: (1)~data-driven methods that learn a constitutive model (i.e., stress-strain relation) from large datasets obtained by solving the microscopic problem for many different inputs, and (2)~projection-based methods that accelerate the microscopic problem by projecting it onto a reduced space. Notable methods for the first class include, for example, the data-driven framework introduced in Kirchdoerfer et al.~\cite{Kirchdoerfer2016} or constitutive artificial neural networks in Linka et al.~\cite{Linka2021}, which were applied to learn effective elastic material models and later extended for other material models (see, e.g.,~\cite{Abdolazizi2023ViscoelasticViscoelasticity,Eggersmann2019,Karapiperis2021}). Other authors generated large datasets and utilized recurrent neural networks to learn history-dependent plasticity material models in, e.g.,~\cite{Mozaffar2019,Wu2022RecurrentStep}. Even though highly accurate and efficient ROMs can be obtained with these methods, there are three concerns: (1)~rather large datasets are usually needed that ideally cover all possible inputs, (2)~dealing with history-dependent behaviors such as plasticity is often challenging, and (3)~extrapolation is unreliable. For the second class, methods based on proper orthogonal decomposition (POD) and hyperreduction have been quite successful. Using POD, the solution space of the microscopic problem is reduced to a fraction of the original problem; with hyperreduction\newtext{~\cite{Chaturantabut2010,Farhat2014DimensionalEfficiency}}, and more specifically, the empirical cubature method (ECM) proposed by Hern\'{a}ndez et al.~\cite{Hernandez2017} and later refined in~\cite{Hernandez2020AECM-hyperreduction}, the assembly of the global stiffness matrix and internal force vector can be performed much more efficiently. Since the microscopic problem is still being solved (in a reduced form), smaller datasets are typically sufficient for good results and dealing with history-dependent behavior is not an issue. In Caicedo et al.~\cite{Caicedo2019HighModeling} and Raschi et al.~\cite{Raschi2021}, two-scale simulations involving highly non-linear RVEs were successfully accelerated. In our previous work~\cite{Guo2023ADeformations}, we constructed an efficient ROM by combining POD and ECM with geometrical transformations that can be applied for two-scale shape optimization problems involving elasto-plastic RVEs under large deformations.

While there is a wide range of works on accelerating the first-order CH model, to the best of our knowledge there have been no attempts towards the reduced-order modeling of any enriched formulations. Inspired by the literature on first-order CH, in this contribution we propose a ROM for second-order CH, utilizing POD and a novel hyperreduction method that is inspired by ECM~\cite{Hernandez2017}. The main reasons for opting for a projection-based ROM over a data-driven one are two-fold: (1)~estimation of parameter bounds for strain gradients is generally difficult a priori, and (2)~the number of input parameters is large, in 2D corresponding to four components of the deformation gradient and another six of the gradient of the deformation gradient (nine and eighteen in 3D). This parameter space does not yet account for any other additional design variables, which would further increase the number of parameters. The main innovative contributions of this manuscript are:
\begin{itemize}
    \item design of a novel hyperreduction algorithm tailored to second-order CH,
    \item development of a hyperreduced POD model for a family of geometrically parameterized microstructures,
    \item derivation of effective quantities arising in reduced second-order CH, and
    \item an empirical analysis of the hyperreduced model for geometrically parameterized two-scale simulations under large deformations and multiscale buckling.
\end{itemize}
After reviewing the theory on second-order CH and specifying the employed formulation used in~\cref{sec2}, the proposed ROM, including the novel hyperreduction algorithm, is presented in~\cref{sec3}. To validate the ROM, numerical examples are discussed in~\cref{sec4} and obtained results are compared with reference solutions in terms of accuracy and efficiency. A summary on the findings with final remarks is provided in~\cref{sec5}.

Throughout the paper, the following notation conventions are used:
\begin{multicols}{2}
\begin{itemize}
    \item scalars $a$,
    \item vectors $\bm{a}=a_i \bm{e}_i$,
    \item position vector $\bm{x}=x_i\bm{e}_i$,
    \item second-order tensors $\bm{A}=A_{ij}\bm{e}_i\bm{e}_j$,
    \item third-order tensors $\bm{\mathcal{A}}=\mathcal{A}_{ijk}\bm{e}_i\bm{e}_j\bm{e}_k$,
    \item fourth-order tensors $\bm{\mathfrak{A}}=\mathfrak{A}_{ijkl}\bm{e}_i\bm{e}_j\bm{e}_k\bm{e}_l$,
    \item matrices $\mathbf{A}$ and column matrices $\mathbf{a}$,
    \item $\bm{a}\cdot\bm{b}=a_i b_i$,
    \item $\bm{a}\otimes\bm{b}=a_i b_j \bm{e}_i\bm{e}_j$,
    \item $\bm{A}\cdot\bm{b}=A_{ij} b_j \bm{e}_i$,
    \item $\bm{A}\cdot\bm{B}=A_{ik} B_{kj} \bm{e}_i\bm{e}_j$,
    \item $\bm{A}:\bm{B}=A_{ij}B_{ji}$,
    \item $\bm{A}\otimes \bm{b}=A_{ij}b_k\bm{e}_i\bm{e}_j\bm{e}_k$,
    \item $\bm{b}\otimes \bm{A}=b_iA_{jk}\bm{e}_i\bm{e}_j\bm{e}_k$,
    \item $\bm{A}\otimes\bm{B}=A_{ij} B_{kl}\bm{e}_i\bm{e}_j\bm{e}_k\bm{e}_l$,
    \item $\bm{a}\cdot\bm{\mathcal{A}}=a_i\mathcal{A}_{ijk} \bm{e}_i\bm{e}_j$,
    \item $\bm{\mathcal{A}}\dotthird\bm{\mathcal{B}} = \mathcal{A}_{ijk}\mathcal{B}_{kji}$,
    \item $\bm{\mathfrak{A}} : \bm{B} = \mathfrak{A}_{ijkl}B_{lk}\bm{e}_i\bm{e}_j$,
    \item $\bm{A} : \bm{\mathfrak{B}} = A_{lk}\mathfrak{B}_{klij}\bm{e}_i\bm{e}_j$,
    \item transpose $\bm{A}^T$, $A_{ij}^T=A_{ji}$, $\bm{\mathcal{A}}^T$, $\mathcal{A}_{ijk}^T=\mathcal{A}_{kji}$,
    \item $\trace{\bm{A}} = A_{ii}$,
    \item gradient operator with respect to $\bm{x}$ \\ $\bm{\nabla}_{\bm{x}} \bm{a}=\dfrac{\partial a_j}{\partial x_i}\bm{e}_i\bm{e}_j$, $\bm{\nabla}_{\bm{x}} \bm{A}=\dfrac{\partial A_{jk}}{\partial x_i}\bm{e}_i\bm{e}_j\bm{e}_k$,
    \item gradient operator with respect to second-order tensor $\bm{B}$ \\
    $\bm{\nabla}_{\bm{B}} \bm{A}=\dfrac{\partial A_{kl}}{\partial B_{ij}}\bm{e}_i\bm{e}_j\bm{e}_k\bm{e}_l$, $\bm{\nabla}_{\bm{B}} \bm{\mathcal{A}}=\dfrac{\partial \mathcal{A}_{klm}}{\partial B_{ij}}\bm{e}_i\bm{e}_j\bm{e}_k\bm{e}_l\bm{e}_m$,
    \item gradient operator with respect to third-order tensor $\bm{\mathcal{B}}$ \\
    $\bm{\nabla}_{\bm{\mathcal{B}}} \bm{A}=\dfrac{\partial A_{lm}}{\partial \mathcal{B}_{ijk}}\bm{e}_i\bm{e}_j\bm{e}_k\bm{e}_l\bm{e}_m$, $\bm{\nabla}_{\bm{\mathcal{B}}} \bm{\mathcal{A}}=\dfrac{\partial \mathcal{A}_{lmn}}{\partial \mathcal{B}_{ijk}}\bm{e}_i\bm{e}_j\bm{e}_k\bm{e}_l\bm{e}_m\bm{e}_n$,
    \item divergence operator with respect to $\bm{x}$ \\ $\bm{\nabla}_{\bm{x}} \cdot \bm{A} = \dfrac{\partial A_{ij}}{\partial x_i}\bm{e}_j$, $\bm{\nabla}_{\bm{x}} \cdot \bm{\mathcal{A}} = \dfrac{\partial A_{ijk}}{\partial x_i}\bm{e}_j\bm{e}_k$,
    \item linearization of functional $\Pi$ around state $\bm{a}$ in direction $\Delta\bm{a}$ \\ $\left.D\Pi\right|_{\bm{a}}\cdot(\Delta\bm{a})=\left.\dfrac{d}{d\tau}\Pi(\bm{a}+\tau\Delta\bm{a})\right|_{\tau=0}$,
\end{itemize}
\end{multicols}
\noindent where the Einstein summation convention is assumed on repeated indices $i,j,k,l,m,n$ and $\bm{e}_i$, $i=1,\dots,d$ denote the basis vectors of a $d$-dimensional Cartesian coordinate frame. Overlines are used to distinguish macroscopic from microscopic quantities.

\section{Second-Order Computational Homogenization}\label{sec2}
\begin{figure}[th]
    \centering
    \includegraphics[width=0.5\textwidth]{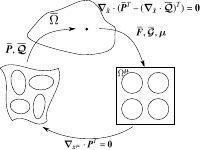}
    \caption{Two-scale formulation in second-order computational homogenization. At every macroscopic point, the deformation gradient $\Fmacro$ and its gradient $\macrobm{\mathcal{G}}$ are used to prescribe boundary conditions for the microscopic problem which, after solving, returns an effective stress $\macrobm{P}$ and higher-order stress $\macrobm{Q}$. The parameter $\bm{\mu}$ describes the geometry of the RVE. More information on $\bm{\mu}$ are provided in~\cref{subsec:second_order_micro}.}
    \label{fig:ch2_twoscale}
\end{figure}

The second-order computational homogenization (CH2) formulation contains the second gradient of the displacement field, thus introducing a length-scale associated with the length-scale of the underlying unit cell, making it possible to capture size and non-local effects~\cite{Mindlin1968OnElasticity,Toupin1962ElasticCouple-stresses}. The formulation of the micro- and macroscopic problem as well as their scale coupling employed in this work is discussed in the subsections below. A schematic sketch of the two-scale problem is depicted in~\cref{fig:ch2_twoscale}.

\subsection{Macroscopic Problem}
In CH2, the macroscopic problem is based on a strain gradient formulation~\cite{Mindlin1968OnElasticity,Toupin1962ElasticCouple-stresses} to model non-local effects of the microstructure. Consider a body $\macro{\Omega} \subset \mathbb{R}^d$ with outer boundaries $\partial\macro{\Omega}$, $d=2,3$ the space dimension, and a position vector $\Xmacro \in \macro{\Omega}$. The governing partial differential equation (PDE) has the form (with body forces neglected for brevity)~\cite{Kouznetsova2004},
\begin{align}
    \derivmacro \cdot (\macrobm{P}(\Fmacro,\macrobm{\mathcal{G}}, \bm{\mu})^T - (\derivmacro \cdot \macrobm{\mathcal{Q}}(\Fmacro,\macrobm{\mathcal{G}}, \bm{\mu}) )^T) = \bm{0}, \label{eq:macro_ch2}
\end{align}
where $\Fmacro \coloneqq \bm{I} + (\derivmacro{\macrobm{u}})^T$ is the macroscopic deformation gradient with $\macrobm{u}(\Xmacro)$ being the macroscopic displacement field and $\bm{I}$ the \newtext{second-order} identity tensor, $\macrobm{\mathcal{G}}\coloneqq\derivmacro{\Fmacro}$ is the gradient of the deformation gradient, i.e., a third-order tensor with symmetry $\macro{\mathcal{G}}_{ijk}=\macro{\mathcal{G}}_{kji}$, and \newtext{$\bm{\mu}$ contains additional parameters}; $\macrobm{P}$ denotes the second-order first Piola-Kirchhoff (1PK) stress tensor, and $\macrobm{\mathcal{Q}}$ is a third-order tensor often referred to as the higher-order or double stress tensor~\cite{Mindlin1968OnElasticity,Toupin1962ElasticCouple-stresses}. By multiplying~\cref{eq:macro_ch2} with a test function $\delta{\macrobm{u}}$ and utilizing the divergence theorem, the following weak form can be derived~\cite{Kouznetsova2004ComputationalMaterials},
\begin{align}
    \integratemacro{\left(\macrobm{P}(\Fmacro,\macrobm{\mathcal{G}}, \bm{\mu}) : \delta\Fmacro^T + \macrobm{\mathcal{Q}}(\Fmacro,\macrobm{\mathcal{G}}, \bm{\mu}) \dotthird \delta\macrobm{\mathcal{G}}\right)} = 
    \int_{\partial \macro{\Omega}_{\text{N}}} \left(\delta\macrobm{u} \cdot (\macrobm{P}-\bm{\nabla}_{\Xmacro} \cdot \macrobm{\mathcal{Q}}) \cdot \macro{\bm{n}} + \macro{\bm{n}} \cdot \macrobm{\mathcal{Q}} : \delta\Fmacro^T\right) d\Xmacro,\label{eq:macro_problem_full}
\end{align}
where $\delta\Fmacro \coloneqq (\derivmacro{\delta\macrobm{u}})^T$ and $\delta\macrobm{\mathcal{G}}\coloneqq\derivmacro{\delta\Fmacro}$ are introduced, $\partial\macro{\Omega}_{\text{N}}$ denotes the boundaries with prescribed Neumann boundary conditions, and $\macro{\bm{n}}$ is the outward unit normal vector. In addition, there are Dirichlet boundaries $\partial\macro{\Omega}_{\text{D}}$, with $\partial\macro{\Omega}_{\text{D}}\cap\partial\macro{\Omega}_{\text{N}}=\emptyset$ and $\partial\macro{\Omega}_{\text{D}}\cup\partial\macro{\Omega}_{\text{N}}=\partial\macro{\Omega}$, where values of the displacement $\macrobm{u}$ and its gradient $\derivmacro{\macrobm{u}}$ are prescribed. In this work, we consider only Dirichlet boundary conditions, so that the terms on the right hand side vanish and~\cref{eq:macro_problem_full} becomes
\begin{align}
    \integratemacro{\left(\macrobm{P}(\Fmacro,\macrobm{\mathcal{G}}, \bm{\mu}) : \delta\Fmacro^T + \macrobm{\mathcal{Q}}(\Fmacro,\macrobm{\mathcal{G}}, \bm{\mu}) \dotthird \delta\macrobm{\mathcal{G}}\right)} \stackrel{!}{=}
    0.\label{eq:macro_problem}
\end{align}
\newtext{The macroscopic problem can then be stated as follows: find $\macrobm{u}$ that fulfills~\cref{eq:macro_problem} for all test functions $\delta\macrobm{u}$.} The relation between $(\macrobm{P},\macrobm{\mathcal{Q}})$ and $(\Fmacro,\macrobm{\mathcal{G}}, \bm{\mu})$ is established by solving the microscopic boundary value problem, which is defined on a representative volume element (RVE) and discussed in more detail in~\cref{subsec:second_order_micro}. In particular, we assume that $\bm{\mu}$ contains parameters that describe the geometry of the RVE. 

In order to solve the problem in~\cref{eq:macro_problem}, the second gradient of $\macrobm{u}$ is required. To this end, Lesicar et al.~\cite{Lesicar2014ADiscretization} employed 36 degrees of freedom (DOF) $C^1$-triangular elements. Wu et al.~\cite{Wu2023Second-orderMetamaterials} utilized an enriched discontinuous Galerkin method combined with a penalty method to enforce $C^1$-continuity weakly. Other works reformulate the problem in~\cref{eq:macro_problem} with a mixed formulation instead~\cite{Kouznetsova2004, Luscher2010AMaterials, RodriguesLopes2022AnHomogenisation}, which \newtext{results in a saddle point problem and} is used also in this work. The idea of the mixed formulation is to introduce an independent deformation gradient field $\widehat{\Fmacro}$ which is coupled with the deformation gradient computed from the displacement field through Lagrange multipliers $\macrobm{L}$. With $\widehat{\Fmacro}$, its gradient $\Gmacro\coloneqq\derivmacro{\widehat{\Fmacro}{}}$, and $\macrobm{L}$, the problem in~\cref{eq:macro_problem} can be rewritten as
\begin{align}
    \macro{\Pi}(\macrobm{u},\widehat{\Fmacro},\macrobm{L}) \coloneqq \integratemacro{\left(\macrobm{P}(\Fmacro, \Gmacro, \bm{\mu}) : \delta\Fmacro^T + \macrobm{\mathcal{Q}}(\Fmacro, \Gmacro, \bm{\mu}) \dotthird \delta\Gmacro + \delta(\macrobm{L} : (\widehat{\Fmacro}{}^T - \Fmacro^T))\right)}\stackrel{!}{=}0, \label{eq:macro_problem_mixed}
\end{align}
\newtext{which should hold for all test functions $\delta\macrobm{u}$, $\delta\widehat{\Fmacro}$ and $\delta\macrobm{L}$.}
Inserting $\delta\Fmacro = (\derivmacro{\delta\macrobm{u}})^T$, $\delta\Gmacro = \derivmacro{\delta\widehat{\Fmacro}}$ and
\begin{align}
    \delta(\macrobm{L} : (\widehat{\Fmacro}{}^T - \Fmacro^T)) &= (\widehat{\Fmacro}{}^T-\Fmacro^T) : \delta\macrobm{L} + \macrobm{L} : \left(\delta\widehat{\Fmacro}{}^T - \derivmacro{\delta\macrobm{u}}\right)
\end{align}
into~\cref{eq:macro_problem_mixed} yields
\begin{align}
\begin{aligned}
    \macro{\Pi}(\macrobm{u},\widehat{\Fmacro},\macrobm{L})&= \integratemacro{\left(\macrobm{P} : \derivmacro{\delta\macrobm{u}} + \macrobm{\mathcal{Q}} \dotthird \derivmacro{\delta\widehat{\Fmacro}} + (\widehat{\Fmacro}{}^T-\Fmacro^T) : \delta\macrobm{L} + \macrobm{L} : \left(\delta\widehat{\Fmacro}{}^T - \derivmacro{\delta\macrobm{u}}\right)\right)} \\
    &= \integratemacro{\left((\macrobm{P} - \macrobm{L}) : \derivmacro{\delta\macrobm{u}} + \macrobm{L} : \delta\widehat{\Fmacro}{}^T + \macrobm{\mathcal{Q}} \dotthird \derivmacro{\delta\widehat{\Fmacro}} + (\widehat{\Fmacro}{}^T-\Fmacro^T) : \delta\macrobm{L}\right)}, \label{eq:macro_problem_mixed_expanded}
\end{aligned}
\end{align}
where arguments of $\macrobm{P}$ and $\macrobm{\mathcal{Q}}$ have been omitted for brevity. Linearization of~\cref{eq:macro_problem_mixed_expanded}, required by the macroscopic iterative Newton solver, around a state $(\macrobm{u},\widehat{\Fmacro},\macrobm{L})$ in directions $(\Delta\macrobm{u},\bm{0},\bm{0})$, $(\bm{0},\Delta\widehat{\Fmacro},\bm{0})$ and $(\bm{0}, \bm{0}, \Delta\macrobm{L})$, yields
\begin{align}
    \left.D\macro{\Pi}\right|_{\macrobm{u},\widehat{\Fmacro},\macrobm{L}} \cdot (\Delta\macrobm{u},\bm{0},\bm{0}) &= \integratemacro{\left(
         \derivmacro{\Delta\macrobm{u}} : \derivmacro[\Fmacro]{\macrobm{P}} : \derivmacro{\delta\macrobm{u}}
        +  \derivmacro{\Delta\macrobm{u}} : \derivmacro[\Fmacro]{\macrobm{\mathcal{Q}}} \dotthird \derivmacro{\delta\widehat{\Fmacro}}
        - \derivmacro{\Delta\macrobm{u}} : \delta\macrobm{L} \right)
    }, \label{eq:linearization_1}\\
    \left.D\macro{\Pi}\right|_{\macrobm{u},\widehat{\Fmacro},\macrobm{L}} \cdot (\bm{0},\Delta\widehat{\Fmacro},\bm{0}) &= \integratemacro{\left(
         \derivmacro{\Delta\widehat{\Fmacro}} \dotthird \derivmacro[\Gmacro]{\macrobm{P}} : \derivmacro{\delta\macrobm{u}}
        +  \derivmacro{\Delta\widehat{\Fmacro}} \dotthird \derivmacro[\Gmacro]{\macrobm{\mathcal{Q}}} \dotthird \derivmacro{\delta\widehat{\Fmacro}}
        + \Delta\widehat{\Fmacro}{}^T : \delta\macrobm{L} \right)
    }, \label{eq:linearization_2}\\
    \left.D\macro{\Pi}\right|_{\macrobm{u},\widehat{\Fmacro},\macrobm{L}} \cdot (\bm{0}, \bm{0}, \Delta\macrobm{L}) &= \integratemacro{\left(
        - \Delta\macrobm{L} : \derivmacro{\delta\macrobm{u}} 
        + \Delta\macrobm{L}  : \delta\widehat{\Fmacro}{}^T    \right)
    }, \label{eq:linearization_3}
\end{align}
where $\derivmacro[\Fmacro]{\macrobm{P}}$, $\derivmacro[\Fmacro]{\macrobm{\mathcal{Q}}}$, $\derivmacro[\Gmacro]{\macrobm{P}}$, and $\derivmacro[\Gmacro]{\macrobm{\mathcal{Q}}}$ are the macroscopic tangents evaluated at $(\Fmacro(\macrobm{u}),\Gmacro(\widehat{\Fmacro}),\bm{\mu})$. Given a suitable discretization for $\macrobm{u}$, $\widehat{\Fmacro}$ and $\macrobm{L}$, the system of~\cref{eq:macro_problem_mixed_expanded,eq:linearization_1,eq:linearization_2,eq:linearization_3} can be solved with the finite element method, when a constitutive relation between $(\macrobm{P},\macrobm{\mathcal{Q}})$ and $(\Fmacro,\Gmacro, \bm{\mu})$ is established. Different combinations of displacement, deformation gradient and Lagrange multiplier shape functions were considered and tested in Kouznetsova et al.~\cite{Kouznetsova2004}. For the numerical examples in this work, quadrilateral elements with eight displacement nodes, four deformation gradient nodes and one Lagrange multiplier node per element are chosen.

\subsection{Parameterized Microscopic Problem}\label{subsec:second_order_micro}
To evaluate $\macrobm{P}$ and $\macrobm{\mathcal{Q}}$ and their derivatives in~\cref{eq:macro_problem_mixed_expanded,eq:linearization_1,eq:linearization_2,eq:linearization_3}, the microscopic problem needs to be solved at every macroscopic integration point. Here, we follow the formulation as presented in Kouznetsova et al.~\cite{Kouznetsova2002Multi-scaleScheme}, where the microscopic problem is modelled as a standard Cauchy continuum. For brevity, a fixed macroscopic material point is assumed, and the dependence on the macroscopic coordinates is omitted in the definition of the microscopic problem provided below.

\begin{figure}[th]
    \centering
    \includegraphics[width=0.7\textwidth]{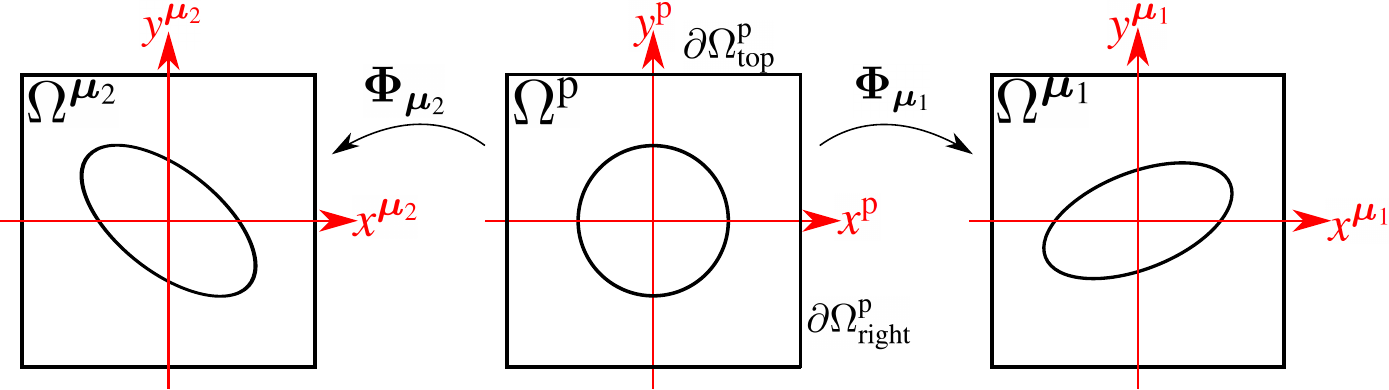}
    \caption{Family of RVE domains $\Omegamu$ parameterized through parameters $\bm{\mu}$. A parent domain $\Omegap$ can be defined which can be transformed through transformations $\Phimu$ into any of the RVE domains $\Omegamu$.}
    \label{fig:ch2_geom}
\end{figure}

Consider a family of domains $\Omegamu \subset \mathbb{R}^{d}$, parameterized by parameters $\bm{\mu}\in\mathbb{P}$ with parameter space $\mathbb{P}$ and spanned by position vectors $\Xmu\in\Omegamu$, see~\cref{fig:ch2_geom}. For all $\bm{\mu}$, the outer boundaries and topology of $\Omegamu$ are assumed to remain fixed. As a consequence, the volume $\left|\Omegamu\right|$ remains constant for all $\bm{\mu}$. Additionally, it is assumed that there exists a parent domain $\Omegap\coloneqq\Omega^{\bm{\mu}^{\text{p}}}$ with $\bm{\mu}^{\text{p}}\in\mathbb{P}$, which can be transformed into any $\Omegamu$ with a transformation map $\Phimu : \Omegap \rightarrow \Omegamu, \Xp \mapsto \Xmu$, transformation gradient $\Fmu \coloneqq (\bm{\nabla}_{\Xp}{\Phimu})^T$ and $d\Xmu = \detFmu d\Xp$. \newtext{Note that the transformation map $\Phimu$ must be one-to-one and thus the absolute value in $\detFmu$ is not strictly necessary, but kept for completeness.} For a fixed domain, i.e., $\Omegap=\Omegamu$, the transformations $\Phimu$ are identity maps, with $\Fmu=\bm{I}$ and $\detFmu=1$. \newtext{Throughout this manuscript, the superscript p will be used to denote quantities related to the parent domain.}

To obtain such transformation maps $\Phimu$, we solve the auxiliary problem as proposed in~\cite{Guo2022LearningParameterizations}. The key idea of the method is to pose an auxiliary linear elasticity problem on the parent domain that can be solved for the transformation displacement $\bm{d}$, which can then be utilized to compute the transformation $\Phimu$ with $\Phimu(\Xp) = \bm{I} + \bm{d}(\Xp)$. The displacement $\bm{d}$ is fixed on the outer boundaries with zero Dirichlet boundary conditions and prescribed on parts of the domain that are known from the parameterization (as an example, the circular interface of $\Omegap$ in~\cref{fig:ch2_geom} is deformed into the elliptical interfaces in $\Omega^{\bm{\mu}_1}$ and $\Omega^{\bm{\mu}_2}$). Subsequently, the auxiliary problem can be solved to find the entire field $\bm{d}(\Xp)$. \newtext{To fully specify the auxiliary problem, a Young's modulus $E^{\text{aux}}$ and Poisson's ratio $\xi^{\text{aux}}$ need to be selected, see~\cite{Guo2022LearningParameterizations}. For all RVEs considered in this work, the auxiliary problem is solved with $E^{\text{aux}}=\SI{1}{MPa}$ and $\xi^{\text{aux}}=0.25$. More details on the auxiliary problem can be found in~\cite{Guo2022LearningParameterizations}}.

The microscopic displacement field $\bm{u}(\Xmu)$ is assumed to consist of a mean field $\macrobm{u}(\Xmu)$ and a fluctuation field $\bm{w}(\Xmu)$, i.e., $\bm{u}(\Xmu) = \macrobm{u}(\Xmu) + \bm{w}(\Xmu)$. The mean field $\macrobm{u}$ is fully prescribed through the macroscopic quantities $(\Fmacro, \Gmacro)$ with
\begin{align}
    \macrobm{u}(\Xmu) \coloneqq (\Fmacro-\bm{I})\cdot\Xmu + \frac{1}{2} (\Xmu\cdot\Gmacro) \cdot \Xmu.
\end{align} 
Subsequently, the microscopic deformation gradient can be defined as
\begin{align}
    \bm{F} \coloneqq \bm{I} + (\bm{\nabla}_{\Xmu}{\bm{u}})^T = \Fmacro + \Xmu \cdot \Gmacro + (\bm{\nabla}_{\Xmu}{\bm{w}})^T.
\end{align}
The governing microscopic PDE is given by,
\begin{align}
    \bm{\nabla}_{\Xmu} \cdot \bm{P}^T(\bm{F}) = \bm{0},
\end{align}
which can be written into its weak form by multiplying with a test function $\delta\bm{w}$ and applying the divergence theorem,
\begin{align}
\begin{aligned}
    \Pi(\bm{w}) &= \integrate[\bm{\mu}]{\bm{\nabla}_{\Xmu}{\delta\bm{w}} : \bm{P}\left(\Fmacro + \Xmu \cdot \Gmacro + (\bm{\nabla}_{\Xmu}{\bm{w}})^T\right)} \stackrel{!}{=} 0,
\end{aligned} \label{eq:weak_form_micro}
\end{align}
where $\bm{P}$ is the microscopic second-order 1PK stress tensor, and the macroscopic quantities $(\Fmacro, \Gmacro)$ act as external forcing terms. For now, no constitutive model at the microscale level is specified, but it is assumed that $\bm{P}$ is a non-linear function of the deformation gradient $\bm{F}$.
The microscopic problem can thus be stated as follows: given $(\Fmacro,\Gmacro,\bm{\mu})$, find $\bm{w}$ that fulfills~\cref{eq:weak_form_micro} for all $\delta\bm{w}$. To remove the dependence of the integral on parameters $\bm{\mu}$,~\cref{eq:weak_form_micro} can be transformed to the parent domain with the transformation map $\Phimu$, i.e.,
\begin{align}
    \Pi(\wp) &= \integrate{\left(\inverseT{\Fmu}\cdot\left(\bm{\nabla}_{\Xp}{\delta\wp}\right)\right) : \Pp(\bm{F}^{\rm{p}})\detFmu} \stackrel{!}{=} 0, \label{eq:weak_form_micro_parent}
\end{align}
where $\wp(\Xp) \coloneqq (\bm{w}\circ\Phimu)(\Xp) = \bm{w}(\Xmu)$, $\delta\wp(\Xp) \coloneqq \delta\bm{w}(\Xmu)$, $\Pp(\Xp)\coloneqq\bm{P}(\Xmu)$, and
\begin{align}
    \bm{F}^{\rm{p}} = \Fmacro + \Phimu(\Xp) \cdot \Gmacro + \left(\bm{\nabla}_{\Xp}{\wp}\right)^T\cdot\inverse{\Fmu}.\label{eq:Fp_ch2}
\end{align}
Hereafter, we write $\Xmu(\Xp)$ instead of $\Phimu(\Xp)$ for brevity. To find the $\wp$ that fulfills~\cref{eq:weak_form_micro_parent} for all $\delta\wp$, the linearization of~\cref{eq:weak_form_micro_parent} around a state $\wp$ in direction $\Delta\wp$ is required,
\begin{align}
    \left.D\Pi\right|_{\wp}\cdot (\Delta\wp) &=  \integrate{\left(\inverseT{\Fmu}\cdot\left(\bm{\nabla}_{\Xp}{\Delta\wp}\right)\right) : \Ap(\bm{F}^{\rm{p}}) : \left(\inverseT{\Fmu}\cdot\left(\bm{\nabla}_{\Xp}{\delta\wp}\right)\right) \detFmu}, \label{eq:weak_form_micro_parent_linearization}
\end{align}
where $\Ap\coloneqq\derivmacro[\bm{F}]{\Pp}$ is the fourth-order stiffness tensor on the parent domain.

To ensure a proper scale transition of the kinematic quantities, different authors derived and proposed additional constraints on the fluctuation field $\wp$. In Kouznetsova et al.~\cite{Kouznetsova2004}, periodic boundary conditions (PBC) for $\wp$ are assumed and the following constraints are derived for a rectangular RVE
\begin{align}
    \int_{\partial \Omegap_{\text{top}}} \wp d\Xp &= \bm{0}, \label{eq:constraint_topedge}\\
    \int_{\partial \Omegap_{\text{right}}} \wp d\Xp &= \bm{0}, \label{eq:constraint_rightedge}
\end{align}
where $\partial \Omegap_{\text{top}}$ and $\partial \Omegap_{\text{right}}$ denote the top and right edge of the RVE $\Omegap$, see~\cref{fig:ch2_geom}. Due to PBC, the same conditions hold for the bottom and left edge. \newtext{In addition, the fluctuation field is set to zero on all four corners of the RVE, i.e., $\wp=\bm{0}$}.
Since it is assumed that $\Omegamu$ has fixed outer boundaries for all $\bm{\mu}$, constraints in~\cref{eq:constraint_topedge,eq:constraint_rightedge} are independent of $\bm{\mu}$. In subsequent works, other authors developed slightly different formulations (see, e.g.,~\cite{Blanco2016TheModel,Luscher2010AMaterials,RodriguesLopes2022AnHomogenisation,Wu2023Second-orderMetamaterials,Yvonnet2020ComputationalBehavior}). In~\cite{RodriguesLopes2022AnHomogenisation}, the authors compared different formulations and pointed out that fixing the corners in the formulation in Kouznetsova et al.~\cite{Kouznetsova2004} leads to stress concentrations and artificial effects at the corners. Instead of fixing the corners, other formulations introduce an additional equation that constrains the rigid body motion with
\begin{align}
    \integrate[\bm{\mu}]{\bm{w}} = \integrate{\wp \detFmu} = \bm{0}. \label{eq:constraint_rigidbody}
\end{align}
The complete microscopic model employed in this work consists of~\cref{eq:weak_form_micro_parent,eq:Fp_ch2,eq:weak_form_micro_parent_linearization,eq:constraint_topedge,eq:constraint_rightedge,eq:constraint_rigidbody} together with PBC for $\wp$. Lagrange multipliers are used to enforce the constraints in~\cref{eq:constraint_topedge,eq:constraint_rightedge,eq:constraint_rigidbody} and PBC, resulting in a saddle point problem.

To solve the microscopic problem, the fluctuation displacement is typically discretized with finite elements (FE). We thus approximate
\begin{align}
    \wp(\Xp) \approx \mathbf{N}(\Xp) \mathbf{w},
\end{align}
where $\mathbf{N}(\Xp)\in\mathbb{R}^{d\times \mathcal{N}}$ denotes the FE shape functions, $\mathbf{w}\in\mathbb{R}^\mathcal{N}$ the coefficients of the discretized displacement fluctuation field, and $\mathcal{N}$ the total number of DOFs. \newtext{Note that lowercase non-italic symbols, e.g., $\mathbf{w}$, are used for column matrices, while uppercase non-italic symbols, e.g., $\mathbf{N}$, are used for matrices.} Subsequently, the weak form in~\cref{eq:weak_form_micro_parent}, together with the constraints, can be written as
\begin{alignat}{2}
\begin{aligned}
    &\mathbf{f}(\mathbf{w}) + \mathbf{C}^T\mathbf{m} &&= \bm{0}, \\
    &\mathbf{C}\mathbf{w} &&= \bm{0},
\end{aligned}\label{eq:mixed_problem_ch2}
\end{alignat}
where $\mathbf{f}\in\mathbb{R}^{\mathcal{N}}$ is the global internal force column matrix, the constraint matrix $\mathbf{C}\in\mathbb{R}^{N_c\times\mathcal{N}}$ is derived from the constraints in~\cref{eq:constraint_topedge,eq:constraint_rightedge,eq:constraint_rigidbody} and PBC, with $N_c$ the number of constraint equations, and $\mathbf{m}\in\mathbb{R}^{N_c}$ are the corresponding Lagrange multipliers. \newtext{Note that the constraint matrix $\mathbf{C}$ depends on the parameter $\bm{\mu}$ because of~\cref{eq:constraint_rigidbody}. This dependency has been dropped for brevity.} Using Newton's method, the non-linear system of equations in~\cref{eq:mixed_problem_ch2} can be solved for $\mathbf{w}$ and $\mathbf{m}$,
\begin{align}
\begin{aligned}
    \begin{bmatrix}
        \mathbf{K}(\mathbf{w}^k) & \mathbf{C}^T\\
        \mathbf{C} & \bm{0}
    \end{bmatrix} \begin{bmatrix}
        \Delta \mathbf{w} \\ \mathbf{m} 
    \end{bmatrix} &= \begin{bmatrix}
        -\mathbf{f}(\mathbf{w}^k) \\ \bm{0}
    \end{bmatrix},\\
    \mathbf{w}^{k+1} &= \mathbf{w}^{k} + \Delta\mathbf{w},
\end{aligned} \label{eq:newton_method_ch2}
\end{align}
where $\mathbf{K}\in\mathbb{R}^{\mathcal{N}\times\mathcal{N}}$ is the global stiffness matrix computed from~\cref{eq:weak_form_micro_parent_linearization}, \newtext{$k\geq0$ is the Newton iteration number, and $\mathbf{w}^0=\mathbf{0}$ is the starting value.} \cref{eq:newton_method_ch2} is repeated until $\left\Vert\mathbf{f}(\mathbf{w}^k) + \mathbf{C}^T\mathbf{m}\right\Vert_2\leq \varepsilon_{\text{newton}}$ with $\varepsilon_{\text{newton}}$ a user-defined tolerance. For more information on the FE method and discretization of weak forms, we refer to~\cite{Belytschko2014NonlinearStructures}.

\subsection{Effective Quantities}
After the microscopic problem has been solved and a solution $\bm{w}^{*\text{p}}$ obtained, the effective stress $\macrobm{P}$, higher-order stress $\macrobm{\mathcal{Q}}$ and their corresponding derivatives with respect to $\Fmacro$ and $\Gmacro$ must be computed. For conciseness, the following microscopic quantities are introduced:
\begin{align}
    \bm{F}^{*\text{p}} &\coloneqq \Fmacro + \Xmu \cdot \Gmacro + \left(\bm{\nabla}_{\Xp}{\bm{w}^{*\text{p}}}\right)^T\cdot\inverse{\Fmu}, \\
    \bm{P}^{*\text{p}} &\coloneqq \Pp(\bm{F}^{*\text{p}}), \\
    \bm{\mathfrak{A}}^{*\text{p}} &\coloneqq \Ap(\bm{F}^{*\text{p}}),
\end{align}
which correspond to the microstructural deformation gradient, 1PK stress, and related stiffness tensors \newtext{all evaluated at the solution $\bm{w}^{*\text{p}}$ on the parent domain, as indicated by the asterisk *}.
Expressions for the effective stress $\macrobm{P}$ and higher-order stress $\macrobm{\mathcal{Q}}$ were derived in~\cite{Kouznetsova2002Multi-scaleScheme} which, after transformation to the parent domain, yield
\begin{align}
    \macrobm{P} &\coloneqq \meanintegrate{\bm{P}^{*\text{p}}\detFmu}, \label{eq:eff_P}\\
    \macrobm{\mathcal{Q}} &\coloneqq \meanintegrate{\frac{1}{2} \left({\bm{P}^{*\text{p}}}^T \otimes \Xmu + \Xmu \otimes \bm{P}^{*\text{p}} \right) \detFmu}. \label{eq:eff_Q}
\end{align}
The effective stiffness components, derived by differentiating the above stress and higher-order stress quantities then yield (in index notation)
\begin{align}
    \deriv[\macro{F}_{kl}]{\macro{P}_{ij}} &= \meanintegrate{\deriv[\macro{F}_{kl}]{P_{ij}^{*\text{p}}}\detFmu}, \label{eq:eff_stiffness_1}\\
    \deriv[\widehat{\macro{\mathcal{G}}}_{mno}]{\macro{P}_{ij}} &= \meanintegrate{\deriv[\widehat{\macro{\mathcal{G}}}_{mno}]{P_{ij}^{*\text{p}}}\detFmu}, \label{eq:eff_stiffness_2}\\
    \deriv[\macro{F}_{mn}]{\macro{\mathcal{Q}}_{ijk}} &= \meanintegrate{\frac{1}{2}\left(\deriv[\macro{F}_{mn}]{P_{ji}^{*\text{p}}}x^{\bm{\mu}}_k + x^{\bm{\mu}}_i \deriv[\macro{F}_{mn}]{P_{jk}^{*\text{p}}}\right)\detFmu}, \label{eq:eff_stiffness_3}\\
    \deriv[\widehat{\macro{\mathcal{G}}}_{mno}]{\macro{\mathcal{Q}}_{ijk}} &= \meanintegrate{\frac{1}{2}\left(\deriv[\widehat{\macro{\mathcal{G}}}_{mno}]{P_{ji}^{*\text{p}}}x^{\bm{\mu}}_k + x^{\bm{\mu}}_i \deriv[\widehat{\macro{\mathcal{G}}}_{mno}]{P_{jk}^{*\text{p}}}\right)\detFmu}.\label{eq:eff_stiffness_4}
\end{align}
In the above,
\begin{align}
    \deriv[\macro{F}_{kl}]{P_{ij}^{*\text{p}}} &= \mathfrak{A}^{*\text{p}}_{ijmn}\left(\delta_{mk}\delta_{nl} + \deriv[\macro{F}_{kl}]{}\left(\deriv[x^{\text{p}}_r]{w^{*\text{p}}_m}\right)\inverse{F_{\bm{\mu},rn}}\right), \label{eq:eff_stiffness_5}
\end{align}
and
\begin{align}
\begin{aligned}
    \deriv[\widehat{\macro{\mathcal{G}}}_{mno}]{P_{ij}^{*\text{p}}} &= \mathfrak{A}^{*\text{p}}_{ijkl}\left(\deriv[\widehat{\macro{\mathcal{G}}}_{mno}]{\left(x^{\bm{\mu}}_r \widehat{\macro{\mathcal{G}}}_{rkl} \right)} + \deriv[\widehat{\macro{\mathcal{G}}}_{mno}]{}\left(\deriv[x^{\text{p}}_s]{w^{*\text{p}}_k}\right)\inverse{F_{\bm{\mu},sl}}\right) \\
    &= \mathfrak{A}^{*\text{p}}_{ijkl}\left(x^{\bm{\mu}}_r \delta_{rm}\delta_{kn}\delta_{lo} + \deriv[\widehat{\macro{\mathcal{G}}}_{mno}]{}\left(\deriv[x^{\text{p}}_s]{w^{*\text{p}}_k}\right)\inverse{F_{\bm{\mu},sl}}\right) \\
    &= \mathfrak{A}^{*\text{p}}_{ijkl}\left(x^{\bm{\mu}}_m\delta_{kn}\delta_{lo} + \deriv[\widehat{\macro{\mathcal{G}}}_{mno}]{}\left(\deriv[x^{\text{p}}_s]{w^{*\text{p}}_k}\right)\inverse{F_{\bm{\mu},sl}}\right).
\end{aligned} \label{eq:eff_stiffness_6}
\end{align}
To determine $\deriv[\macro{F}_{kl}]{\vphantom{1}}\left(\deriv[x^{\text{p}}_r]{w^{*\text{p}}_m}\right)$ and $\deriv[\widehat{\macro{\mathcal{G}}}_{mno}]{\vphantom{1}}\left(\deriv[x^{\text{p}}_s]{w^{*\text{p}}_k}\right)$,~\cref{eq:weak_form_micro_parent} is differentiated with respect to $\Fmacro$ and $\Gmacro$ to derive linear tangent problems that can be solved to find the corresponding sensitivities. As an example, for one particular component $\macro{F}_{kl}$ (where the indices $k$ and $l$ are assumed to be temporarily fixed), the differentiation yields
\begin{align}
    \deriv[\macro{F}_{kl}]{\Pi(\bm{w}^{*\text{p}})} = &\integrate{\left(\inverseT{\Fmu}\cdot\left(\bm{\nabla}_{\Xp}{\delta\wp}\right)\right) : \deriv[\macro{F}_{kl}]{\bm{P}^{*\text{p}}} \detFmu} = 0,
\end{align}
which can be rearranged with~\cref{eq:eff_stiffness_5} as
\begin{align}
\begin{aligned}
&\integrate{\left(\inverseT{\Fmu}\cdot\left(\bm{\nabla}_{\Xp}{\bm{q}_{kl}}\right)\right) : \bm{\mathfrak{A}}^{*\text{p}} : \left(\inverseT{\Fmu}\cdot\left(\bm{\nabla}_{\Xp}{\delta\wp}\right)\right) \detFmu} \\ &= -\bm{E}_{kl}^T : \left(\integrate{\bm{\mathfrak{A}}^{*\text{p}} : \left(\inverseT{\Fmu}\cdot\left(\bm{\nabla}_{\Xp}{\delta\wp}\right)\right) \detFmu}\right),  
\end{aligned}\label{eq:tangentproblem_F}
\end{align}
where a new auxiliary vector field $\bm{q}_{kl} \coloneqq \deriv[\macro{F}_{kl}] {\bm{w}^{*\text{p}}}$ is defined, reflecting the sensitivity of the microfluctuation field with respect to the change of the macroscopic deformation gradient, and $\bm{E}_{kl} \in \mathbb{R}^{d\times d}$ is a second-order tensor with all entries zero except for the $kl$-th entry which is 1. The linear problem of~\cref{eq:tangentproblem_F} is solved for all combinations $k,l=1,\dots,d$ to obtain $\bm{q}_{kl}$ for each component of $\Fmacro$. The same procedure is followed for $\Gmacro$. With an auxiliary vector field $\bm{q}_{mno} \coloneqq \deriv[\widehat{\macro{\mathcal{G}}}_{mno}]{\bm{w}^{*\text{p}}}$, the differentiation of~\cref{eq:weak_form_micro_parent} for one particular component $\widehat{\macro{\mathcal{G}}}_{mno}$ (where the indices $m$, $n$ and $o$ are assumed to be temporarily fixed) yields, together with~\cref{eq:eff_stiffness_6},
\begin{align}
\begin{aligned}
    &\integrate{\left(\inverseT{\Fmu}\cdot\left(\bm{\nabla}_{\Xp}{\bm{q}_{mno}}\right)\right) : \bm{\mathfrak{A}}^{*\text{p}} : \left(\inverseT{\Fmu}\cdot\left(\bm{\nabla}_{\Xp}{\delta\wp}\right)\right) \detFmu} \\ &= -\bm{\mathcal{E}}_{mno}^T \dotthird \left(\integrate{\Xmu \otimes \bm{\mathfrak{A}}^{*\text{p}} : \left(\inverseT{\Fmu}\cdot\left(\bm{\nabla}_{\Xp}{\delta\wp}\right)\right) \detFmu}\right), 
\end{aligned} \label{eq:tangentproblem_G}
\end{align}
where $\bm{\mathcal{E}}_{mno}\in \mathbb{R}^{d\times d\times d}$ is a third-order tensor with all entries zero except for the $mno$-th entry which is 1. The linear problem of~\cref{eq:tangentproblem_G} is then solved for all combinations $m,n,o=1,\dots,d$ to obtain $\bm{q}_{mno}$ for each component of $\Gmacro$.

\section{Reduced-Order Modeling}\label{sec3}
Since the microscopic problem is solved at every macroscopic integration point, its solution must be efficient. Due to the often complicated RVE geometries, a fine discretization (using, e.g., finite elements) is required, resulting in a large number of DOFs and integration points, which entail a costly solution. Furthermore, computing the effective quantities for a fine RVE mesh presents another computationally expensive operation. To construct a reduced-order model (ROM) for a more efficient solution, we employ two reduction techniques: (1)~we utilize proper orthogonal decomposition (POD) to reduce the number of DOFs in~\cref{sec:pod}; and (2)~to establish a more efficient integration scheme, we propose in~\cref{sec:hyperecm} a novel hyperreduction algorithm that uses ideas of the empirical cubature method~\cite{Hernandez2017}, which is specifically suited for the second-order CH formulation.

\subsection{Proper Orthogonal Decomposition}\label{sec:pod}
To reduce the number of DOFs, the fluctuation displacement field $\wp$ is approximated with a reduced basis~\cite{Hesthaven2016,Quarteroni2015}, i.e.,
\begin{align}
    \wp \approx \sum_{n=1}^N a_n \bm{v}_n, \label{eq:rb_w}
\end{align}
where $N$ is typically much smaller than $\mathcal{N}$, i.e., $N\ll\mathcal{N}$. The global basis functions, $\{\bm{v}_n\}_{n=1}^N$, are obtained by applying POD to a set of pre-computed snapshots of $\wp$ for different parameter values $(\Fmacro,\Gmacro,\bm{\mu})$. Since each of the basis functions is computed from a linear combination of pre-computed periodic solutions that fulfill the constraints in~\cref{eq:constraint_topedge,eq:constraint_rightedge,eq:constraint_rigidbody}, every basis function is periodic and also fulfills~\cref{eq:constraint_rightedge,eq:constraint_topedge}. This implies that any solution $\wp$ that is represented by~\cref{eq:rb_w} always fulfills these conditions. However, the basis functions will only fulfill the constraint in~\cref{eq:constraint_rigidbody} if a fixed geometry is assumed for the RVE, i.e., $\bm{\mu}$ is constant. For varying geometries, the constraint is violated due to the influence of $\detFmu$. Nevertheless, in our numerical examples, tests with and without enforcing~\cref{eq:constraint_rigidbody} through Lagrange multipliers were run and only insignificant differences of the solutions were observed. For that reason, we do not enforce~\cref{eq:constraint_rigidbody} for the ROM. This has the added advantage that no constraints have to be considered for the ROM and, thus, more efficient solvers for the resulting system of linear equations can be utilized.

\newtext{\begin{rem}
For the full problem,~\cref{eq:constraint_rigidbody} must be enforced to remove rigid body modes and ensure a unique solution. However, for the ROM, each basis function is a linear combination of snapshots that fulfill~\cref{eq:constraint_rigidbody} in their respective domains. Such basis functions should usually not span a solution space that contains a rigid body mode of the microscopic problem for any new domain. If numerical issues are encountered during the solution without the constraint, one should enforce the constraint.
\end{rem}}

By inserting~\cref{eq:rb_w} into~\cref{eq:weak_form_micro_parent,eq:weak_form_micro_parent_linearization} and assuming a Galerkin projection, the components of the internal force $\mathbf{f}\in\mathbb{R}^N$ and global stiffness matrix $\mathbf{K}\in\mathbb{R}^{N\times N}$ can be computed:
\begin{align}
    f_i(\mathbf{a}) &\coloneqq \integrate{\left(\inverseT{\Fmu}\cdot\left(\bm{\nabla}_{\Xp}\bm{v}_i\right)\right) : \Pp(\bm{F}^{\rm{p}}) \detFmu}, \label{eq:forcevector_pod}\\
    K_{ij}(\mathbf{a}) &\coloneqq \integrate{\left(\inverseT{\Fmu}\cdot\left(\bm{\nabla}_{\Xp}\bm{v}_j\right)\right) : \Ap(\bm{F}^{\rm{p}}) : \left(\inverseT{\Fmu}\cdot\left(\bm{\nabla}_{\Xp}\bm{v}_i\right)\right) \detFmu}, \label{eq:stiffnessmatrix_pod}
\end{align}
for all $i, j=1,\dots, N$\newtext{, and
\begin{align}
    \bm{F}^{\rm{p}}(\mathbf{a}) &= \Fmacro + \Xmu \cdot \Gmacro + \left(\sum_{n=1}^N a_n (\bm{\nabla}_{\Xp}\bm{v}_n)^T\right) \cdot \inverse{\Fmu}.
\end{align}
} The column matrix $\mathbf{a}=[a_1,\dots,a_N]^T$ contains the unknown coefficients to be solved for.

\subsection{Hyperreduction}\label{sec:hyperecm}
While the reduced system of~\cref{eq:forcevector_pod,eq:stiffnessmatrix_pod} only has $N$ DOFs, computing the integrals in~\cref{eq:forcevector_pod,eq:stiffnessmatrix_pod} (as well as~\cref{eq:eff_P,eq:eff_Q,eq:eff_stiffness_1,eq:eff_stiffness_2,eq:eff_stiffness_3,eq:eff_stiffness_4,eq:eff_stiffness_5,eq:eff_stiffness_6,eq:tangentproblem_F,eq:tangentproblem_G} for the effective quantities) requires integration over the whole full finite element mesh (typically with many Gauss quadrature points). To accelerate this computation, a more efficient integration scheme (i.e., fewer integration points and corresponding weights) is sought that closely approximates the numerical integration with Gaussian quadrature of the following four quantities:
\begin{itemize}
    \item Internal force $\mathbf{f}=[f_1,\dots,f_N]^T$ in~\cref{eq:forcevector_pod}:
    \begin{align}
    \begin{aligned}
        f_i &= \integrate{\left(\inverseT{\Fmu}\cdot\left(\bm{\nabla}_{\Xp}{\bm{v}_i}\right) \right) : \Pp \detFmu} \\
        &= \integrate{\bm{\nabla}_{\Xp}{\bm{v}_i} : \Wp}, 
    \end{aligned}\label{eq:integral_f}
    \end{align}
    for all $i=1,\dots,N$ and where the weighted stress $\Wp\coloneqq \Pp \cdot \Fmu^{-T}\detFmu$ is defined.

    \item Effective stress $\macrobm{P}$ in~\cref{eq:eff_P}:
    \begin{align}
        \begin{aligned}
            \macrobm{P} &= \meanintegrate{\Pp \detFmu} \\
                        &= \meanintegrate{\Wp\cdot\Fmu^T} \\
                        &= \meanintegrate{\Wp},
        \end{aligned}\label{eq:integral_W}
    \end{align}
    where, in the last line the invariance of the integral with respect to $\Fmu^T$ was used, which was proven in~\cite[Appendix A]{Guo2022LearningParameterizations}. This implies that the accurate integration of the effective stress is equivalent to the accurate integration of the weighted stress $\Wp$.

    \item Effective higher-order stress $\macrobm{\mathcal{Q}}$ in~\cref{eq:eff_Q}:
    \begin{align}
    \begin{aligned}
        \macrobm{\mathcal{Q}} &= \meanintegrate{\frac{1}{2} \left({\Pp}^T \otimes \Xmu + \Xmu \otimes \Pp \right) \detFmu} \\
        &= \meanintegrate{\bm{\mathcal{Y}}^{\text{p}}},
    \end{aligned} \label{eq:integral_Q}
    \end{align}
    where the weighted higher-order stress $\bm{\mathcal{Y}}^{\text{p}}\coloneqq \dfrac{1}{2} \left({\Pp}^T \otimes \Xmu + \Xmu \otimes \Pp \right) \detFmu$ is defined.

    \item Volume:
    \begin{align}
        V \coloneqq \left|\Omegap\right| = \integrate{}. \label{eq:integral_volume}
    \end{align}
\end{itemize}
Even though the integration of the volume does not necessarily have to be accurate, it helps to stabilize the algorithm used to find the new integration points. In particular, it leads to fewer weights that are equal to 0 during the solution of the non-negative least squares problem introduced later in~\cref{eq:wls_ecm}.

\subsubsection{Algorithm}
To find an efficient integration scheme, we use concepts from the empirical cubature method (ECM), which was previously applied to identify integration points and weights for an efficient integration of the internal force, as proposed by Hern\'{a}ndez et al.~\cite{Hernandez2017} and recently extended to varying geometries in Guo et al.~\cite{Guo2023ADeformations}. In the first step, similarly to the fluctuation field $\wp$, snapshots of the weighted stress $\Wp$ and weighted higher-order stress $\bm{\mathcal{Y}}^{\text{p}}$ are collected for different parameter values $(\Fmacro,\Gmacro,\bm{\mu})$. Utilizing POD, two sets of basis functions for $\Wp$ and $\bm{\mathcal{Y}}^{\text{p}}$, $\{\bm{B}_m\}_{m=1}^M$ and $\{\bm{\mathcal{H}}_l\}_{l=1}^L$, are found, with which $\Wp$ and $\bm{\mathcal{Y}}^{\text{p}}$ can be approximated, i.e.,
\begin{align}
    \Wp &\approx \sum_{m=1}^M \alpha_m \bm{B}_m, \label{eq:rb_W}\\
    \bm{\mathcal{Y}}^{\text{p}} &\approx \sum_{l=1}^L \beta_l \bm{\mathcal{H}}_l. \label{eq:rb_Y}
\end{align}
Inserting~\cref{eq:rb_W,eq:rb_Y} into~\cref{eq:integral_f,eq:integral_W,eq:integral_Q} yields
\begin{align}
    f_i &\approx \sum_{m=1}^M \alpha_m \integrate{\bm{\nabla}_{\Xp}{\bm{v}_i} : \bm{B}_m}, \quad \forall i=1,\dots,N, \label{eq:integral_f_approx}\\
    \macrobm{P} &\approx \sum_{m=1}^M \alpha_m \meanintegrate{\bm{B}_m}, \label{eq:integral_W_approx}\\
    \macrobm{\mathcal{Q}} &\approx \sum_{l=1}^L \beta_l \meanintegrate{\bm{\mathcal{H}}_l}. \label{eq:integral_Q_approx}
\end{align}
Since~\cref{eq:integral_f_approx,eq:integral_W_approx,eq:integral_Q_approx} should be accurately integrated for any values of $\alpha_m$ and $\beta_l$, all the occurring integrals in the sums should be approximated accurately. Together with the volume equation in~\cref{eq:integral_volume}, integration points and weights are sought that approximate these $NM + d^2M + d^3L + 1$ integrals accurately. The factors $d^2$ and $d^3$ arise due to the number of components of $\macrobm{P}$ and $\macrobm{\mathcal{Q}}$.

Consider for now the full set of Gaussian integration points $\{\widehat{\bm{x}}_q, \newtext{\widehat{h}_q}\}_{q=1}^{\widehat{Q}}$ corresponding to the fully resolved discretization, where $\widehat{Q}$ is the total number of Gauss integration points, $\widehat{\bm{x}}_q$ their positions and $\newtext{\widehat{h}_q}$ their weights. By defining 
\begin{align}
    \bm{1} \coloneqq \begin{bmatrix}
        1, \dots, 1
    \end{bmatrix}^T \in \mathbb{R}^{\widehat{Q}},
\end{align}
and the flattened basis functions (in 2D, i.e., $d=2$),
\begin{align}
    \underline{\bm{B}}_m &\coloneqq \begin{bmatrix}
        B_{m,11}, B_{m,12}, B_{m,21}, B_{m,22}
    \end{bmatrix}^T\in\mathbb{R}^{d^2}, \quad \forall m=1,\dots,M, \\
    \underline{\bm{\mathcal{H}}}_l &\coloneqq \begin{bmatrix}
        \mathcal{H}_{l,111}, \mathcal{H}_{l,112}, \mathcal{H}_{l,121}, \mathcal{H}_{l,122}, \mathcal{H}_{l,211}, \mathcal{H}_{l,212}, \mathcal{H}_{l,221}, \mathcal{H}_{l,222}
    \end{bmatrix}^T\in\mathbb{R}^{d^3}, \quad \forall l=1,\dots,L,
\end{align}
the numerical approximation of the $NM + d^2M + d^3L + 1$ integrals with the full Gauss quadrature can be written in algebraic form
\begin{align}
    \underbrace{\begin{bmatrix}
        \mathbf{A}_1 \\ \mathbf{A}_2 \\ \mathbf{A}_3 \\ \bm{1}^T
    \end{bmatrix}}_{\eqqcolon \mathbf{A}}
    \underbrace{\begin{bmatrix}
        \newtext{\widehat{h}_1} \\ \vdots \\ \newtext{\widehat{h}_{\widehat{Q}}}
    \end{bmatrix}}_{\eqqcolon \newtext{\widehat{\mathbf{h}}}}
    = \underbrace{\begin{bmatrix}
        \mathbf{b}_1 \\ \mathbf{b}_2 \\ \mathbf{b}_3 \\ V
    \end{bmatrix}}_{\eqqcolon \mathbf{b}}, \label{eq:full_integration}
\end{align}
where
\newtext{\begin{align*}
    \mathbf{A}_1 = \begin{bmatrix} \left.\left(\bm{\nabla}_{\Xp}{\bm{v}_1} : \bm{B}_1\right)\right|_{\widehat{\bm{x}}_1} & \hdots & \left.\left(\bm{\nabla}_{\Xp}{\bm{v}_1} : \bm{B}_1\right)\right|_{\widehat{\bm{x}}_{\widehat{Q}}} \\ 
    \vdots & & \vdots \\
    \left.\left(\bm{\nabla}_{\Xp}{\bm{v}_1} : \bm{B}_M\right)\right|_{\widehat{\bm{x}}_1} & \hdots & \left.\left(\bm{\nabla}_{\Xp}{\bm{v}_1} : \bm{B}_M\right)\right|_{\widehat{\bm{x}}_{\widehat{Q}}} \\ 
    \vdots & & \vdots \\ 
    \left.\left(\bm{\nabla}_{\Xp}{\bm{v}_N} : \bm{B}_1\right)\right|_{\widehat{\bm{x}}_1} & \hdots & \left.\left(\bm{\nabla}_{\Xp}{\bm{v}_N} : \bm{B}_1\right)\right|_{\widehat{\bm{x}}_{\widehat{Q}}} \\ 
    \vdots & & \vdots \\
    \left.\left(\bm{\nabla}_{\Xp}{\bm{v}_N} : \bm{B}_M\right)\right|_{\widehat{\bm{x}}_1} & \hdots & \left.\left(\bm{\nabla}_{\Xp}{\bm{v}_N} : \bm{B}_M\right)\right|_{\widehat{\bm{x}}_{\widehat{Q}}} \end{bmatrix} \in \mathbb{R}^{NM \times \widehat{Q}},
\end{align*}
\begin{align*}
    \mathbf{A}_2 = \begin{bmatrix} \underline{\bm{B}}_1(\widehat{\bm{x}}_1) & \hdots & \underline{\bm{B}}_1(\widehat{\bm{x}}_{\widehat{Q}})\\ 
    \vdots & & \vdots \\ 
    \underline{\bm{B}}_M(\widehat{\bm{x}}_1) & \hdots & \underline{\bm{B}}_M(\widehat{\bm{x}}_{\widehat{Q}})    \end{bmatrix} \in \mathbb{R}^{d^2M \times \widehat{Q}}, \
    \mathbf{A}_3 = \begin{bmatrix} \underline{\bm{\mathcal{H}}}_1(\widehat{\bm{x}}_1) & \hdots & \underline{\bm{\mathcal{H}}}_1(\widehat{\bm{x}}_{\widehat{Q}})\\ 
    \vdots & & \vdots \\ 
    \underline{\bm{\mathcal{H}}}_L(\widehat{\bm{x}}_1) & \hdots & \underline{\bm{\mathcal{H}}}_L(\widehat{\bm{x}}_{\widehat{Q}})    \end{bmatrix} \in \mathbb{R}^{d^3L \times \widehat{Q}},
\end{align*}
and
\begin{align*}
    \mathbf{b}_1 = \begin{bmatrix} \integrate{\bm{\nabla}_{\Xp}{\bm{v}_1} : \bm{B}_1} \\ \vdots \\ \integrate{\bm{\nabla}_{\Xp}{\bm{v}_1} : \bm{B}_M} \\ \vdots \\ \integrate{\bm{\nabla}_{\Xp}{\bm{v}_N} : \bm{B}_1} \\ \vdots \\ \integrate{\bm{\nabla}_{\Xp}{\bm{v}_N} : \bm{B}_M}    \end{bmatrix} \in \mathbb{R}^{NM}, \quad
    \mathbf{b}_2 = \begin{bmatrix} \integrate{\underline{\bm{B}}_1} \\ \vdots \\ \integrate{\underline{\bm{B}}_M}    \end{bmatrix} \in \mathbb{R}^{d^2M}, \quad
    \mathbf{b}_3 = \begin{bmatrix} \integrate{\underline{\bm{\mathcal{H}}}_1} \\ \vdots \\ \integrate{\underline{\bm{\mathcal{H}}}_L}    \end{bmatrix} \in \mathbb{R}^{d^3L}.
\end{align*}}
The system in~\cref{eq:full_integration} can be equivalently rewritten as
\begin{align}
    \underbrace{\begin{bmatrix}
        \widehat{\mathbf{A}}_1 \\ \widehat{\mathbf{A}}_2 \\ \widehat{\mathbf{A}}_3 \\ \bm{1}^T
    \end{bmatrix}}_{\eqqcolon\widehat{\mathbf{A}}}
    \newtext{\widehat{\mathbf{h}}}
    = \underbrace{\begin{bmatrix}
        \mathbf{0} \\ \mathbf{0} \\ \mathbf{0} \\ V
    \end{bmatrix}}_{\eqqcolon \widehat{\mathbf{b}}}, \label{eq:full_integration_reformulated}
\end{align}
with 
\newtext{\begin{align}
    \widehat{\mathbf{A}}_i &= \mathbf{A}_i - \frac{1}{V} \mathbf{b}_i \otimes \bm{1}, \quad i=1,2,3,
\end{align}}
which is convenient for the definition of residuals of the algorithm discussed below.

The goal now is to select a subset of integration points $\{\bm{x}_q\}_{q=1}^Q$ from the set of all integration points, i.e., $\{\bm{x}_q\}_{q=1}^Q \subset \{\widehat{\bm{x}}_q\}_{q=1}^{\widehat{Q}}$, such that $Q\ll\widehat{Q}$, with corresponding weights $\{\newtext{h_q}\}_{q=1}^Q$ obtained by minimizing the following weighted non-negative least squares residual,
\begin{align}
\begin{aligned}
    \newtext{\mathbf{h}}^{\text{LS}} &= \argmin_{\newtext{\mathbf{h}}\geq\bm{0}} \norm{\widehat{\mathbf{b}} - \widehat{\mathbf{A}}_{\bullet \mathcal{I}} \newtext{\mathbf{h}}}_{\mathbf{\Sigma}} \\ 
    &= \argmin_{\newtext{\mathbf{h}}\geq\bm{0}} \norm{\widehat{\mathbf{A}}\newtext{\widehat{\mathbf{h}}} - \widehat{\mathbf{A}}_{\bullet \mathcal{I}} \newtext{\mathbf{h}}}_{\mathbf{\Sigma}} \\ 
    &= \argmin_{\newtext{\mathbf{h}}\geq\bm{0}} \norm{\widehat{\mathbf{r}}(\newtext{\mathbf{h}})}_{\mathbf{\Sigma}},
\end{aligned} \label{eq:wls_ecm}
\end{align}
where $\Vert\mathbf{a}\Vert_{\mathbf{\Sigma}}\coloneqq \sqrt{\mathbf{a}^T\mathbf{\Sigma}\mathbf{a}}$ and the residual 
\begin{align}
    \widehat{\mathbf{r}}(\newtext{\mathbf{h}})\coloneqq \widehat{\mathbf{A}}\newtext{\widehat{\mathbf{h}}} - \widehat{\mathbf{A}}_{\bullet \mathcal{I}} \newtext{\mathbf{h}} \label{eq:wls_residual}
\end{align}
are defined, $\mathcal{I}$ denotes a set of non-repeating indices with $\left|\mathcal{I}\right|=Q$ and $\widehat{\mathbf{A}}_{\bullet \mathcal{I}}$ is the submatrix of $\widehat{\mathbf{A}}$ with $Q$ selected columns according to the entries of $\mathcal{I}$. The matrix $\bm{\Sigma}$ is a weighting matrix with a block diagonal structure
\begin{align}
    \bm{\Sigma} = \begin{bmatrix}
        c_1\bm{\Sigma}_1 & \bm{0} & \bm{0} & \bm{0} \\
        \bm{0} & c_2\bm{\Sigma}_2 & \bm{0} & \bm{0} \\
        \bm{0} & \bm{0} & c_3\bm{\Sigma}_3 & \bm{0} \\
        \bm{0} & \bm{0} & \bm{0} & \bm{\Sigma}_4 \\
    \end{bmatrix},\label{eq:ecm_weights}
\end{align}
where each block corresponds to one of the approximated quantities defined as,
\begin{alignat}{2}
    \bm{\Sigma_1} &= \diag(\sigma^{\bm{w}}_1\sigma^{\newtext{\bm{Y}}}_1,\hdots,\sigma^{\bm{w}}_1\sigma^{\newtext{\bm{Y}}}_M,\hdots,\sigma^{\bm{w}}_N\sigma^{\newtext{\bm{Y}}}_1,\hdots,\sigma^{\bm{w}}_N\sigma^{\newtext{\bm{Y}}}_M) &&\in \mathbb{R}^{NM \times NM}, \\
    \bm{\Sigma_2} &= \diag(\underbrace{\sigma^{\newtext{\bm{Y}}}_1,\hdots,\sigma^{\newtext{\bm{Y}}}_1}_{\text{$d^2$ times}},\hdots,\underbrace{\sigma^{\newtext{\bm{Y}}}_M,\hdots,\sigma^{\newtext{\bm{Y}}}_M}_{\text{$d^2$ times}}) &&\in \mathbb{R}^{d^2M \times d^2M}, \\
    \bm{\Sigma_3} &= \diag(\underbrace{\sigma^{\bm{\mathcal{Y}}}_1,\hdots,\sigma^{\bm{\mathcal{Y}}}_1}_{\text{$d^3$ times}},\hdots,\underbrace{\sigma^{\bm{\mathcal{Y}}}_L,\hdots,\sigma^{\bm{\mathcal{Y}}}_L}_{\text{$d^3$ times}}) &&\in \mathbb{R}^{d^3L \times d^3L}, \\
    \bm{\Sigma_4} &= \diag(1) &&\in \mathbb{R}^{1 \times 1}.
\end{alignat}
The entries $\sigma_i^{\bm{w}}, \sigma_m^{\newtext{\bm{Y}}}, \sigma_l^{\bm{\mathcal{Y}}}$ for all $i=1,\dots,N$, $m=1,\dots,M$, and $l=1,\dots,L$ correspond to the ordered normalized singular values of the POD of the fluctuation field $\wp$, weighted stress $\Wp$ and weighted higher-order stress $\bm{\mathcal{Y}}^{\text{p}}$, with $\sigma_1^{\bm{w}}=\sigma_1^{\newtext{\bm{Y}}}=\sigma_1^{\bm{\mathcal{Y}}}=1$. The parameters $c_1,c_2,c_3$ enable control over the importance of each of the approximated quantities. Their influence is illustrated in~\cref{sec4}. The blocks of the weighting matrix $\bm{\Sigma}$ are chosen in this fashion to promote the integration scheme to approximate the basis functions corresponding to larger singular values more accurately than the ones corresponding to smaller singular values. The indices in $\mathcal{I}$ are selected one by one, similarly to the greedy algorithm presented in~\cite{Hernandez2017}. The exact algorithm on the selection is provided in~\cref{algo:ecm}. Here, the residual $\widehat{\mathbf{r}}$ is split into four parts, i.e.,
\begin{align}
    \widehat{\mathbf{r}} = \begin{bmatrix}
        \widehat{\mathbf{r}}_1^T, \widehat{\mathbf{r}}_2^T, \widehat{\mathbf{r}}_3^T, \widehat{\mathbf{r}}_4^T
    \end{bmatrix}^T,
\end{align}
where $\widehat{\mathbf{r}}_i$ for $i=1,\dots,4$ are the residuals for each quantity. Independent residuals are introduced to check that each quantity is approximated accurately up to a precision, depending on the choice of tolerances $\varepsilon_1$, $\varepsilon_2$, $\varepsilon_3$, $\varepsilon_4$, with,
\begin{align}
    r_1 \coloneqq \frac{\norm{\widehat{\mathbf{r}}_1}_{\bm{\Sigma}_1}}{\trace{\bm{\Sigma}_1}} < \varepsilon_1, \quad
    r_2 \coloneqq \frac{\norm{\widehat{\mathbf{r}}_2}_{\bm{\Sigma}_2}}{\trace{\bm{\Sigma}_2}} < \varepsilon_2, \quad
    r_3 \coloneqq\frac{\norm{\widehat{\mathbf{r}}_3}_{\bm{\Sigma}_3}}{\trace{\bm{\Sigma}_3}} < \varepsilon_3, \quad
    r_4 \coloneqq\frac{\norm{\widehat{\mathbf{r}}_4}_{\bm{\Sigma}_4}}{V} < \varepsilon_4,
    \label{eq:ecm_tolerances}
\end{align}
where $r_i$ for $i=1,\dots,4$ are the standardized norms of the residuals. As will be shown in~\cref{sec4}, all $r_i$ generally decay with different rates, which can, however, be tuned with the parameters $c_1,c_2,c_3$. The lowest number of quadrature points can be achieved when all residuals reach the desired tolerances at the same time. This will be demonstrated in~\cref{sec4}.

\begin{algorithm}
\caption{Integration point selection algorithm}\label{algo:ecm}
\begin{algorithmic}
\Require $\widehat{\mathbf{A}}, \widehat{\mathbf{b}}, \bm{\Sigma}, \varepsilon_1, \varepsilon_2, \varepsilon_3, \varepsilon_4, k_{\text{max}}$
\Ensure $\newtext{\mathbf{h}}^{\text{LS}}, \mathcal{I}$
\State Initialize empty list of selected columns $\mathcal{I} \gets \emptyset$
\State Initialize list of candidate indices $\mathcal{C} \gets \{1,\dots,\widehat{Q}\}$
\State Set iteration number $k \gets 0$
\State Set initial residual $\widehat{\mathbf{r}} \gets \widehat{\mathbf{b}}$
\While{$k < k_{\text{max}}$}
\State $ k \gets k + 1$
\State Find the column $i$ of $\widehat{\mathbf{A}}$ with $$i=\argmax_{j\in\mathcal{C}} \frac{\widehat{\mathbf{A}}_{\bullet j}^T \mathbf{\Sigma}\widehat{\mathbf{r}}}{\sqrt{\widehat{\mathbf{A}}_{\bullet j}^T \mathbf{\Sigma}\widehat{\mathbf{A}}_{\bullet j}}}$$
\State Add selected index $\mathcal{I} \gets \mathcal{I} \cup \{i\}$
\State Remove selected index from candidates $\mathcal{C} \gets \mathcal{C} \setminus \{i\}$
\State Solve~\cref{eq:wls_ecm} for $\newtext{\mathbf{h}}^{\text{LS}}$
\State Compute residuals $\widehat{\mathbf{r}}_i$ according to~\cref{eq:wls_residual}
\If{all conditions in~\cref{eq:ecm_tolerances} are fulfilled}
    \State \Return $\newtext{\mathbf{h}}^{\text{LS}}$, $\mathcal{I}$  \Comment{Algorithm \newtext{has} converged}
\EndIf
\EndWhile
\end{algorithmic}
\end{algorithm}

\section{Numerical Examples}\label{sec4}
To examine and illustrate different features of the proposed ROM, two macroscopic example problems with a parameterized microstructure are studied in two dimensions and under plane strain conditions. The results are compared against the full two-scale second-order CH solution (later referred to as CH2) as well as the direct numerical simulation (DNS), where the microstructure is fully resolved at the macroscale. The ROM is discussed in detail in the first example, whereas the second example shows a possible application, in which a full DNS might not be feasible anymore (especially in three dimensions), but the ROM computes an excellent approximation in a reasonable amount of time. Note that the examples considered here could potentially also be solved with a satisfactory accuracy using first-order CH (CH1). However, as the focus of this work is on the reduced-order modeling of CH2, no comparisons with CH1 are conducted. A systematic and quantitative comparison of CH1 and CH2 applied to mechanical metamaterials can be found in~\cite{Sperling2024AMetamaterials}.

\begin{figure}[th]
    \centering
    \begin{subfigure}[b]{0.24\textwidth}
        \centering
        \includegraphics[width=\textwidth]{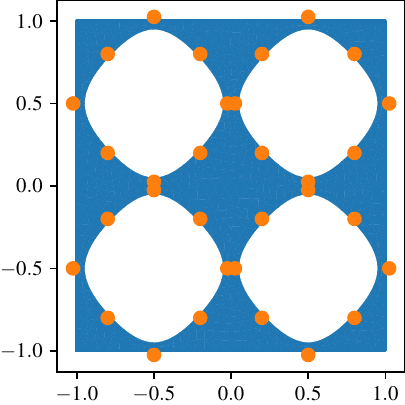}
        \caption{$\zeta=\SI{-0.075}{mm}$}
        \label{fig:rve_definition1}
    \end{subfigure}
    \hfill
    \begin{subfigure}[b]{0.24\textwidth}
        \centering
        \includegraphics[width=\textwidth]{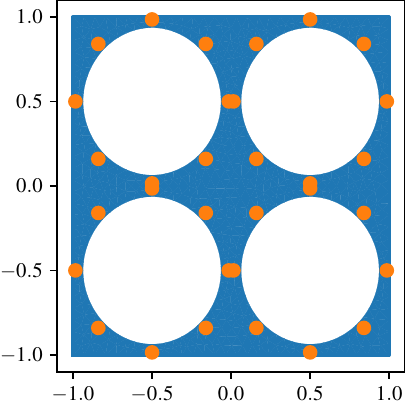}
        \caption{$\zeta=\SI{-0.035}{mm}$}
        \label{fig:rve_definition2}
    \end{subfigure}
    \hfill
    \begin{subfigure}[b]{0.24\textwidth}
        \centering
        \includegraphics[width=\textwidth]{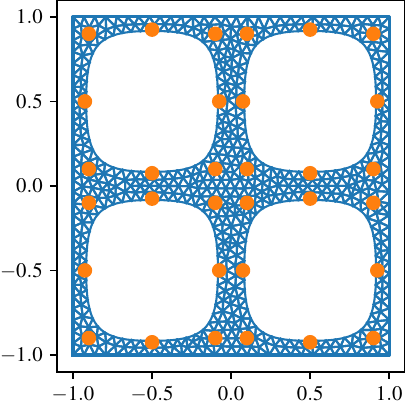}
        \caption{$\zeta=\SI{0.025}{mm}$}
        \label{fig:rve_parent_mesh}
     \end{subfigure}
     \hfill
    \begin{subfigure}[b]{0.24\textwidth}
        \centering
        \includegraphics[width=\textwidth]{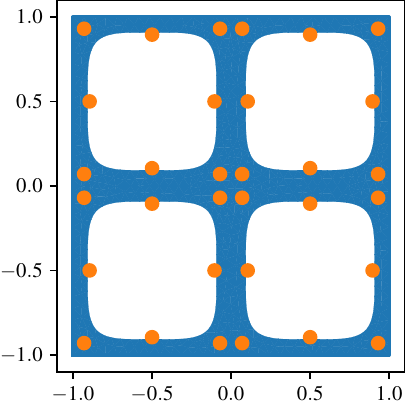}
        \caption{$\zeta=\SI{0.055}{mm}$}
        \label{fig:rve_definition4}
     \end{subfigure}
     \caption{Example geometries for (a) $\zeta=\SI{-0.075}{mm}$, (b) $\zeta=\SI{-0.035}{mm}$, (c) $\zeta=\SI{0.025}{mm}$, and (d) $\zeta=\SI{0.055}{mm}$. The control points defining the hole shapes are shown in orange and the matrix material in blue. Depending on $\zeta$, the shape of the holes is more circular or square-like, and the RVE more prone to local or global buckling. The parent domain $\Omegap$ is chosen for $\zeta=\SI{0.025}{mm}$ with a simulation mesh consisting of 1066 six-noded triangular elements with 4882 DOFs.}
    \label{fig:rve_definition}
\end{figure}

For both examples, a metamaterial-based RVE with four identical holes is selected, motivated by \textit{Specimen 1} in Bertoldi et al.~\cite{Bertoldi2008MechanicsStructures}. The size of the RVE is $\SI{2}{mm}\times\SI{2}{mm}$ and the local coordinate system is chosen in the center of the domain, i.e., the domain of the RVE is given by $[\SI{-1}{mm},\SI{1}{mm}]^2$, see~\cref{fig:rve_definition}. Each hole is described by a cubic B-spline with eight control points, of which the coordinates are parameterized by one geometrical parameter $\bm{\mu}=\{\zeta\}$. For the top right hole, the coordinates (in $\si{mm}$) of the control points are $(0.05 + \zeta, 0.5)$, $(0.125 - \zeta, 0.125 - \zeta)$, $(0.5, 0.05 + \zeta)$, $(0.875 + \zeta, 0.125 - \zeta)$, $(0.95 - \zeta, 0.5)$, $(0.875 + \zeta, 0.875 + \zeta)$, $(0.5, 0.95 - \zeta)$, $(0.125 - \zeta, 0.875 + \zeta)$. The coordinates of the control points for the other holes are obtained by shifting the coordinates of the top right hole by $\SI{1}{mm}$ in the $x$- and/or $y$-direction, and the same $\zeta$ is assumed for each hole. The geometry of the RVE is shown for different values of $\zeta=\{\SI{-0.075}{mm}, \SI{-0.035}{mm}, \SI{0.025}{mm}, \SI{0.055}{mm}\}$ in~\cref{fig:rve_definition}. The parent domain $\Omegap$, chosen with $\zeta=\SI{0.025}{mm}$, and its simulation mesh, consisting of 4882 DOFs and 1066 six-noded triangular elements, are shown in~\cref{fig:rve_parent_mesh}. A mesh convergence study was conducted to ensure that the effective quantities obtained with this mesh are converged with respect to the element size. Note that the control points (in orange color) are allowed to lie outside the RVE domain as long as the resulting B-spline curves do not intersect with the outer boundary of the RVE.

Depending on $\zeta$, the shape of the holes changes from a circular shape to a square-like one. For circular holes, it is known that the RVE buckles locally under compression due to the symmetry and exhibits auxetic behavior~\cite{Bertoldi2008MechanicsStructures}, i.e., under uniaxial compression in one direction the RVE contracts in the perpendicular direction. On the other hand, square-like holes promote global buckling on the macroscale instead of local buckling. This is illustrated in~\cref{subsec:ex1}, where significantly different behaviors of the macrostructure are observed when varying $\zeta$.

The RVE material is modelled as a hyperelastic Mooney--Rivlin material with strain energy density function
\begin{align}
    W(\bm{F}) = C_1(I_1-3) + C_2(I_1 - 3)^{2} - 2C_1\log{J} + \frac{K}{2}(J - 1)^{2}, \label{eq:StrainDensity}
\end{align}
where $I_{1}\coloneqq\trace{\bm{C}}$ is the first invariant of the right Cauchy--Green tensor $\bm{C} \coloneqq \bm{F}^T \bm{F}$ and $J = \det{\bm{F}}$ characterizes the volume change. The constants $C_{1}$, $C_{2}$ and $K$ are material parameters, which are set to $C_{1}=\SI{0.55}{MPa}$, $C_{2}=\SI{0.3}{MPa}$, and $K=\SI{55}{MPa}$, according to the experimental data in Bertoldi et al.~\cite{Bertoldi2008MechanicallyStructures}. 

\subsection{Uniaxial Compression of a Perforated Plate}\label{subsec:ex1}
\begin{figure}[th]
    \centering
    \begin{subfigure}[b]{0.45\textwidth}
        \centering
        \includegraphics[width=\textwidth]{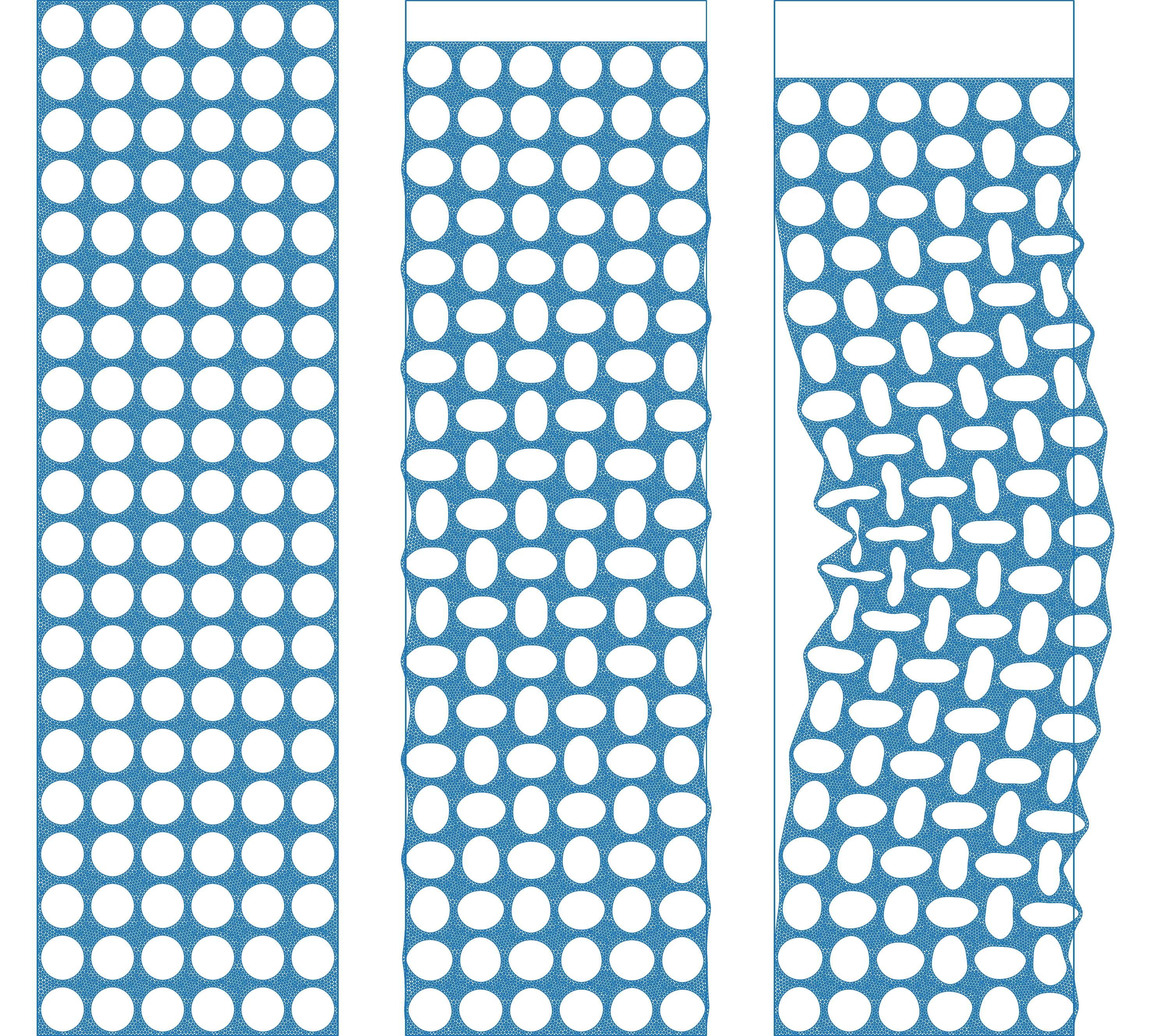}
        \caption{$\zeta=\SI{-0.035}{mm}$}
        \label{fig:plate_dns1}
    \end{subfigure}
    \qquad
    \begin{subfigure}[b]{0.45\textwidth}
        \centering
        \includegraphics[width=\textwidth]{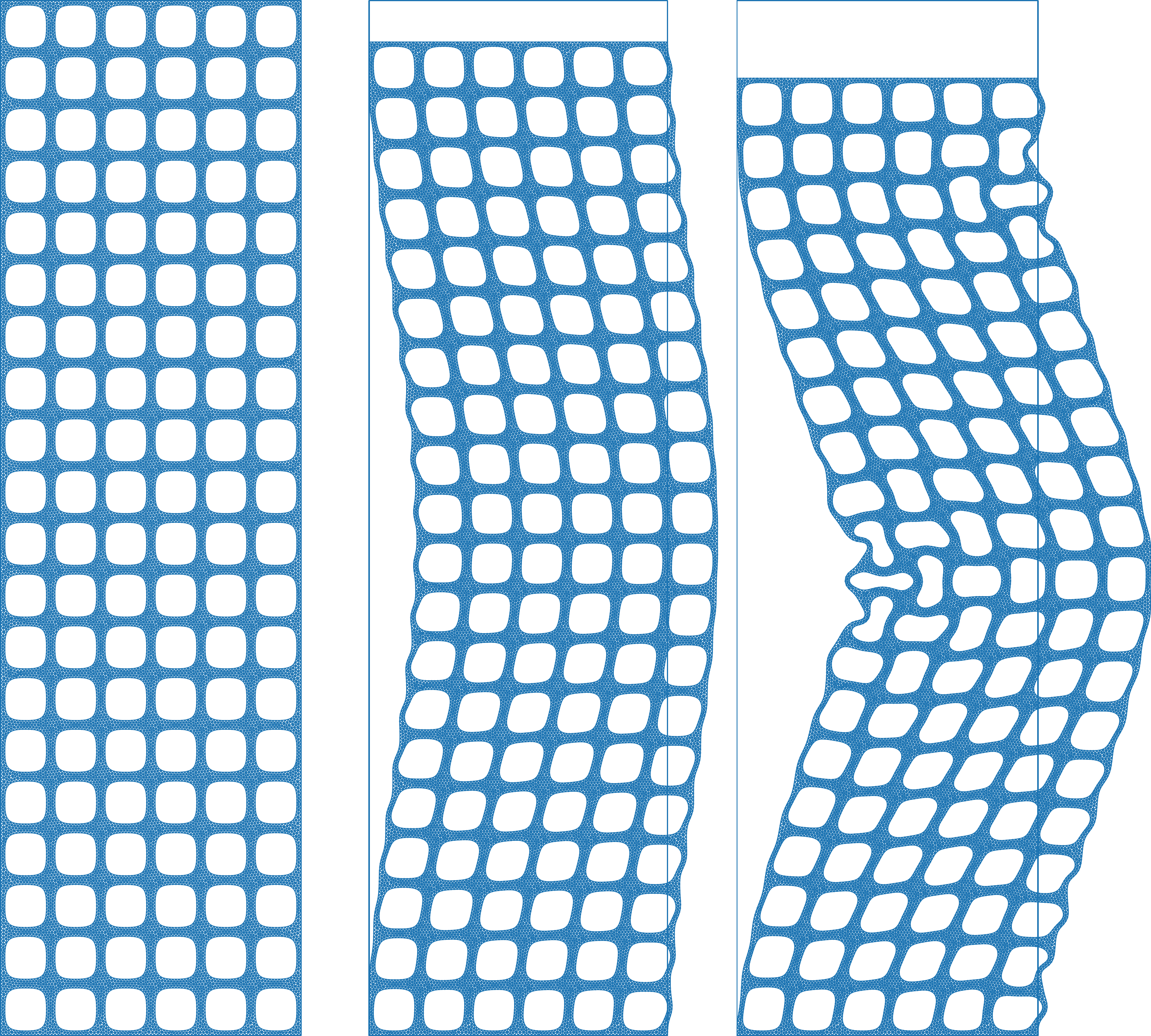}
        \caption{$\zeta=\SI{0.03}{mm}$}
        \label{fig:plate_dns2}
    \end{subfigure}
     \caption{DNS solutions for $\zeta=\SI{-0.035}{mm}$ in (a), $\zeta=\SI{0.03}{mm}$ in (b). In each panel, the undeformed (left) and deformed states at 4\% (middle) and 7.5\% (right) compression are shown. For $\zeta=\SI{-0.035}{mm}$ the structure first buckles locally and then globally, while for $\zeta=\SI{0.03}{mm}$ the structure first buckles globally with subsequent local patterning.}
    \label{fig:plate_dns}
\end{figure}

In the first example, uniaxial compression of a rectangular perforated plate of size $W\times H$ (width $W=\SI{6}{mm}$, height $H=\SI{20}{mm}$) in the longitudinal direction is considered. The top edge is compressed up to $7.5\%$ strain, while the bottom edge is fixed, and the geometrical parameter $\zeta$ is constant throughout the macrostructure. As the reference solution, full DNS solutions, where the microstructure is fully resolved and for which triangular six-noded elements are used, are computed for various values of $\zeta$. For each DNS model, the number of elements is approximately 32,000 and number of DOFs 140,000. For $\zeta=\{\SI{-0.035}{mm}, \SI{0.03}{mm}\}$, the undeformed state together with the deformed solutions at 4\% and 7.5\% strain are shown in~\cref{fig:plate_dns}. For $\zeta=\SI{-0.035}{mm}$ the macrostructure first buckles locally (patterning of holes) and then globally (entire macrostructure buckles). For $\zeta=\SI{0.03}{mm}$ the structure first buckles globally and then locally, implying that the overall behavior of the macrostructure changes significantly for different $\zeta$.

For the homogenized plate, a uniform mesh, consisting of two elements in the horizontal and four elements in the vertical direction with a total of 32 quadrature points, is chosen, amounting to 32 microscopic problems that must be solved for each macroscopic Newton iteration. \newtext{This discretization was found to give converged results in terms of the reaction force.} Regarding the boundary conditions, the displacement is fixed at the bottom edge and prescribed at the top edge with value $\tilde{u}$. \newtext{Since the prescribed displacements do not vary with the $x$-axis, we accordingly set the $xx$- and $yx$-components of the deformation gradient to $\widehat{\macro{F}}_{xx}=1$ and $\widehat{\macro{F}}_{yx}=0$ at the top and bottom edge.} To prevent zero energy modes corresponding to the components $\widehat{\macro{F}}_{xy}$ and $\widehat{\macro{F}}_{yy}$, the deformation gradient at the bottom left point is fully fixed with $\widehat{\Fmacro}=\bm{I}$.

To construct the ROM, training data must be generated by solving the microscopic problem for different input parameters $(\Fmacro,\Gmacro,\bm{\mu})$, which span an 11-dimensional parameter space in 2D. In total, 100 samples are generated randomly for the loading parameters $(\Fmacro,\Gmacro)$ from a uniform distribution with parameter bounds provided in~\cref{table:rve_bounds1}. Since the macrostructure is compressed up to 7.5\% in the $y$-direction and locally higher deformations might occur, the lower bound for $\macro{F}_{yy}-1$ is chosen as $-0.1$. Since the RVE behaves auxetically, the lower bound for $\macro{F}_{xx}-1$ is also assumed to be $-0.1$. The upper bound for both $\macro{F}_{xx}-1$ and $\macro{F}_{yy}-1$ is chosen as $0.02$ to capture some tensile behavior of the RVE. Due to the global buckling, large shear strains might occur and bounds of $[-0.1, 0.1]$ are chosen for $\macro{F}_{xy}$ and $\macro{F}_{yx}$. Bounds for $\Gmacro$ are difficult to estimate without prior knowledge. Here, every component is assumed to range from $\SI{-0.05}{\per\milli\metre}$ to $\SI{0.05}{\per\milli\metre}$, which for the RVE size of $\SI{2}{mm}\times\SI{2}{mm}$ can result in maximal \newtext{components of the deformation gradient/strains} in the range of $[-0.1, 0.1]$ with $\bm{F}-\bm{I} = \Xp \cdot \Gmacro$ and $\Xp\in\Omegap=[\SI{-1}{mm},\SI{1}{mm}]^2$.

Subsequently, all samples are divided into five groups, each with 20 samples and assigned one value of $\zeta=\{\SI{-0.05}{mm}, -\SI{0.025}{mm}, \SI{0.0}{mm}, \SI{0.025}{mm}, \SI{0.05}{mm}\}$. For each sample, the macroscopic loads are applied to the RVE with $(t\Fmacro,t\Gmacro)$, where $t\in[0,1]$ is a parameterization pseudo-time increased linearly from $0$ to $1$ in 20 equidistant load steps, resulting in 20 snapshots per sample. In total, 2000 snapshots are obtained which are all used for the construction of the ROM.

\begin{table}[th]
\centering
\caption{Parameter bounds used for sampling training data for the ROM. The bounds for $\Fmacro$ are motivated by the applied macroscopic compression loads and the auxeticity of the RVE. Bounds for $\Gmacro$ are assumed to range from $\SI{-0.05}{\per\milli\metre}$ to $\SI{0.05}{\per\milli\metre}$.}
\resizebox{0.8\textwidth}{!}{\begin{tabular}{cccccc}
                                                 & \multicolumn{1}{c|}{$\macro{F}_{xx}-1$}            & \multicolumn{1}{c|}{$\macro{F}_{xy}$}            & \multicolumn{1}{c|}{$\macro{F}_{yx}$}            & $\macro{F}_{yy}-1$                                 &                             \\ \cline{2-5}
                                                 & \multicolumn{1}{c|}{$[-0.1, 0.02]$}              & \multicolumn{1}{c|}{$[-0.1, 0.1]$}               & \multicolumn{1}{c|}{$[-0.1, 0.1]$}               & $[-0.1, 0.02]$                                   &                             \\
\multicolumn{1}{l}{}                             & \multicolumn{1}{l}{}                             & \multicolumn{1}{l}{}                             & \multicolumn{1}{l}{}                             & \multicolumn{1}{l}{}                             & \multicolumn{1}{l}{}        \\
\multicolumn{1}{c|}{$\widehat{\macro{\mathcal{G}}}_{xxx}$ $[\si{\per\milli\metre}]$} & \multicolumn{1}{c|}{$\widehat{\macro{\mathcal{G}}}_{xxy}$ $[\si{\per\milli\metre}]$} & \multicolumn{1}{c|}{$\widehat{\macro{\mathcal{G}}}_{xyx}$ $[\si{\per\milli\metre}]$} & \multicolumn{1}{c|}{$\widehat{\macro{\mathcal{G}}}_{xyy}$ $[\si{\per\milli\metre}]$} & \multicolumn{1}{c|}{$\widehat{\macro{\mathcal{G}}}_{yxy}$ $[\si{\per\milli\metre}]$} & $\widehat{\macro{\mathcal{G}}}_{yyy}$ $[\si{\per\milli\metre}]$\\ \hline
\multicolumn{1}{c|}{$[-0.05, 0.05]$}             & \multicolumn{1}{c|}{$[-0.05, 0.05]$}             & \multicolumn{1}{c|}{$[-0.05, 0.05]$}             & \multicolumn{1}{c|}{$[-0.05, 0.05]$}             & \multicolumn{1}{c|}{$[-0.05, 0.05]$}             & $[-0.05, 0.05]$
\end{tabular}}
\label{table:rve_bounds1}
\end{table}

\subsubsection{Results}
The accuracy and efficiency of the ROM depends on several factors:
\begin{itemize}
    \item the number of basis functions for the fluctuation displacement $N$, weighted stress $M$ and weighted higher-order stress $L$,
    \item the error tolerances $\varepsilon_1$, $\varepsilon_2$, $\varepsilon_3$ and $\varepsilon_4$, and
    \item the hyperparameters $c_1$, $c_2$ and $c_3$ that control the weighting matrix $\bm{\Sigma}$ of the weighted least squares problem in~\cref{eq:wls_ecm,eq:ecm_weights}.
\end{itemize}
To choose an adequate amount of basis functions, the singular values of POD are often utilized, as they give an indication on the information loss due to the reduction. Given the ordered singular values $\{\sigma_i\}_{i=1}^{N_{\rm{S}}}$ of POD, a criterion can be defined with:
\begin{align}
    1 - \frac{\sum_{i=1}^{N_{\rm{POD}}} \sigma_i^2}{\sum_{i=1}^{N_{\text{S}}} \sigma_i^2} < \mathcal{E}_{\rm{POD}}, \label{eq:pod_tol}
\end{align}
where $N_{\text{S}}$ denotes the total number of training snapshots and $\mathcal{E}_{\rm{POD}}$ is a user-specified tolerance. The number of basis functions is then selected to be equal to the smallest $N_{\rm{POD}}$, for which~\cref{eq:pod_tol} is fulfilled. For the weighted stress and higher-order stress, good results were obtained with $\mathcal{E}_{\rm{POD}}=\expnumber{5}{-3}$, for which $M=28$ and $L=28$ were found. For the fluctuation field, three values of $\mathcal{E}_{\rm{POD}}=\expnumber{1}{-4}$, $\expnumber{1}{-5}$ and $\expnumber{1}{-6}$ were considered, which resulted in $N=48$, $78$ and $112$ basis functions.

Regarding the error tolerances $\varepsilon_1$, $\varepsilon_2$, $\varepsilon_3$ and $\varepsilon_4$, numerical tests for different values were carried out and a good balance in terms of accuracy and efficiency was found for $\varepsilon_1=\varepsilon_2=\varepsilon_3=\varepsilon_4=\expnumber{1}{-4}$.

\begin{figure}[th]
    \centering
    \begin{subfigure}[b]{0.32\textwidth}
        \centering
        \includegraphics[width=\textwidth]{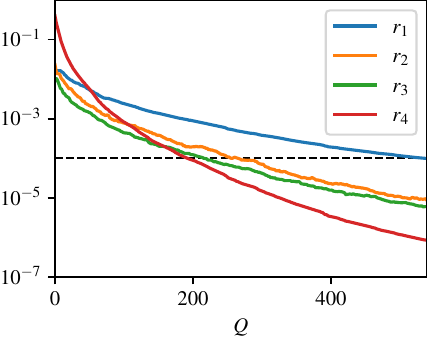}
        \caption{$c_1=1$, $c_2=1$, $c_3=1$}
        \label{fig:resid1}
    \end{subfigure}
    \hfill
    \begin{subfigure}[b]{0.32\textwidth}
        \centering
        \includegraphics[width=\textwidth]{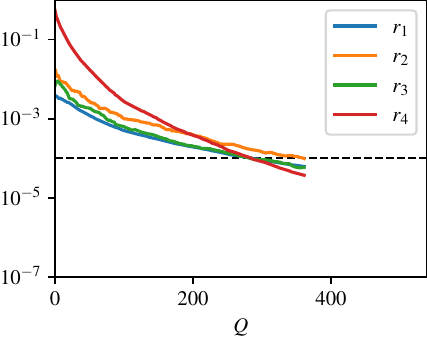}
        \caption{$c_1=10$, $c_2=1$, $c_3=1$}
        \label{fig:resid2}
    \end{subfigure}
    \hfill
    \begin{subfigure}[b]{0.32\textwidth}
        \centering
        \includegraphics[width=\textwidth]{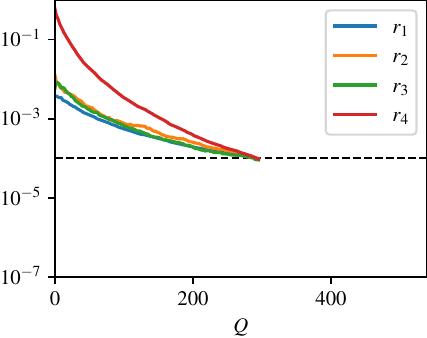}
        \caption{$c_1=10$, $c_2=1.6$, $c_3=1.1$}
        \label{fig:resid3}
    \end{subfigure}
     \caption{Decay of standardized norm of residuals $r_1$, $r_2$, $r_3$ and $r_4$\newtext{, defined in~\cref{eq:ecm_tolerances},} over the number of selected integration points $Q$ with $N=48$, $M=28$, $L=28$ and $\varepsilon_1=\varepsilon_2=\varepsilon_3=\varepsilon_4=\expnumber{1}{-4}$. The hyperparameters $c_1$, $c_2$ and $c_3$ control the rate of decay for each $r_i$, which results in a different number of required quadrature points \newtext{to reach the same level of accuracy}: (a) $Q=543$, (b) $Q=358$, whereas (c) only $Q=297$ quadrature points are selected. \newtext{The same $x$-axis range is used for all plots to highlight that much fewer quadrature points are selected for different combinations of $c_i$.}}
    \label{fig:resid}
\end{figure}

The choice of the hyperparameters $c_1$, $c_2$ and $c_3$ affects the rates with which each of the standardized norm of residuals $r_1$, $r_2$, $r_3$ and $r_4$ (see~\cref{eq:ecm_tolerances} for the definition) decreases over the number of selected quadrature points $Q$. In general, the lowest number of quadrature points can be found when $c_1$, $c_2$ and $c_3$ are tuned such that all $r_i$ fall below their corresponding tolerances at roughly the same time. In~\cref{fig:resid}, the decay of each $r_i$ over the selected number of quadrature points with $N=48$, $M=28$, $L=28$ and $\varepsilon_1=\varepsilon_2=\varepsilon_3=\varepsilon_4=\expnumber{1}{-4}$ is shown for different choices of $c_1$, $c_2$ and $c_3$. For $c_1=c_2=c_3=1$ (see~\cref{fig:resid1}), it can be clearly seen that $r_1$ drops much more slowly than $r_2$, $r_3$ and $r_4$, resulting in a total of $Q=543$ quadrature points. When increasing $c_1$ to $10$ (see~\cref{fig:resid2}), $r_1$ drops more quickly, ending up in a total number of $Q=358$ quadrature points. Finally, for values of $c_1=10$, $c_2=1.6$ and $c_3=1.1$ (see~\cref{fig:resid3}) all tolerances are achieved at roughly the same time with only $Q=297$ quadrature points. For $N=78$ and $112$, $Q=318$ and $337$ quadrature points are found when all other hyperparameters are kept constant.

\begin{figure}[th]
    \centering
    \begin{subfigure}[b]{0.32\textwidth}
        \centering
        \includegraphics[width=\textwidth]{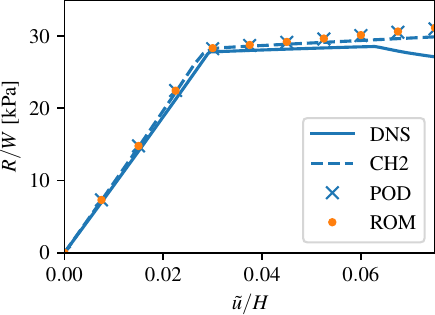}
        \caption{$N=48$, $Q=297$}
        \label{fig:comparison_-0.035_1}
    \end{subfigure}
    \hfill
    \begin{subfigure}[b]{0.32\textwidth}
        \centering
        \includegraphics[width=\textwidth]{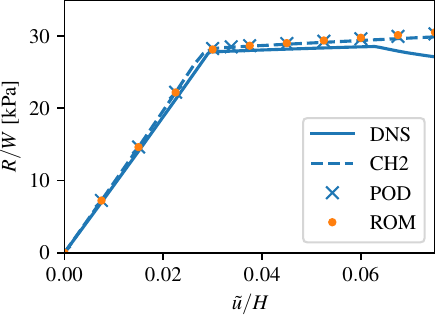}
        \caption{$N=78$, $Q=318$}
        \label{fig:comparison_-0.035_2}
    \end{subfigure}
    \hfill
    \begin{subfigure}[b]{0.32\textwidth}
        \centering
        \includegraphics[width=\textwidth]{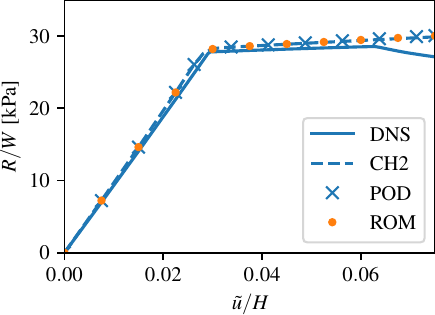}
        \caption{$N=112$, $Q=337$}
        \label{fig:comparison_-0.035_3}
    \end{subfigure}
     \caption{Force-displacement curves for $\zeta=\SI{-0.035}{mm}$ (i.e., circular holes buckling first locally) for different numbers of basis functions $N$ for the fluctuation field. The ROM solution closely follows the POD solution, implying that the proposed hyperreduction algorithm yields accurate results. For increasing number of fluctuation displacement basis functions $N$, the POD and ROM solution both approach that of CH2.}
    \label{fig:comparison_-0.035}
\end{figure}

To evaluate the accuracy of the ROM for $N=48$, $78$ and $112$ basis functions for the displacement, the two-scale compression of the perforated plate is solved for $\zeta=\SI{-0.035}{mm}$ (recall that the training data was sampled for $\zeta=\{\SI{-0.05}{mm}, -\SI{0.025}{mm}, \SI{0.0}{mm}, \SI{0.025}{mm}, \SI{0.05}{mm}\}$), and the total resulting reaction force $R$ acting on the top edge is plotted over the prescribed displacement $\tilde{u}$ and compared to the DNS, CH2 and POD solutions in~\cref{fig:comparison_-0.035} ($R$ and $\tilde{u}$ are normalized with the width $W$ and height $H$ of the plate to yield nominal quantities). Here, POD denotes the solution obtained with the POD basis in~\cref{eq:rb_w}, but with full integration of the reduced system, i.e., computing the integrals in~\cref{eq:forcevector_pod,eq:stiffnessmatrix_pod} (as well as~\cref{eq:eff_P,eq:eff_Q,eq:eff_stiffness_1,eq:eff_stiffness_2,eq:eff_stiffness_3,eq:eff_stiffness_4,eq:eff_stiffness_5,eq:eff_stiffness_6,eq:tangentproblem_F,eq:tangentproblem_G}) with Gauss quadrature. The ROM closely follows the POD solution, showing that the reduced integration is very accurate. It is also clear that the prebuckling stage and buckling point predicted by CH2 are sufficiently accurate already for $N=48$, but small deviations from the CH2 solution can be observed for increasing $\tilde{u}$ in the postbuckling stage. This error is decreased by increasing $N$ to $78$ or $112$. Moreover, it can be observed that CH2 predicts a slightly higher prebuckling stiffness than the DNS, which was also observed in~\cite{Sperling2024AMetamaterials}, and is unable to predict the second (global) buckling point (at around $6\%$ strain), illustrating some modeling limitations of the CH2 scheme.

\begin{figure}[th]
    \centering
    \begin{subfigure}[b]{0.32\textwidth}
        \centering
        \includegraphics[width=\textwidth]{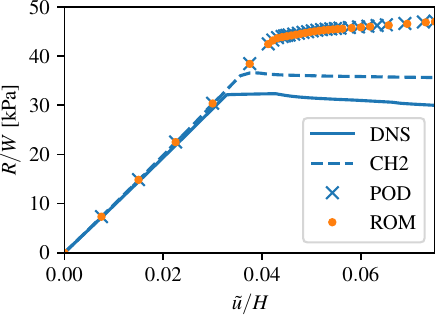}
        \caption{$N=48$, $Q=297$}
        \label{fig:comparison_0.03_1}
    \end{subfigure}
    \hfill
    \begin{subfigure}[b]{0.32\textwidth}
        \centering
        \includegraphics[width=\textwidth]{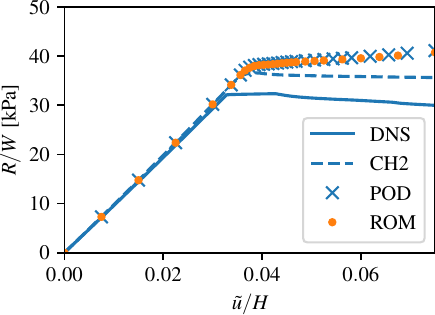}
        \caption{$N=78$, $Q=318$}
        \label{fig:comparison_0.03_2}
    \end{subfigure}
    \hfill
    \begin{subfigure}[b]{0.32\textwidth}
        \centering
        \includegraphics[width=\textwidth]{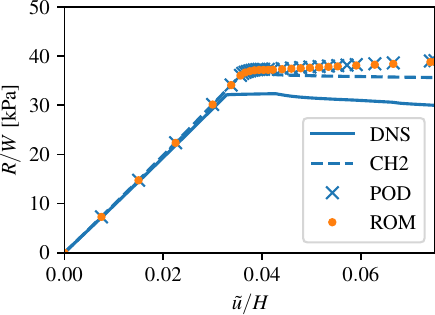}
        \caption{$N=112$, $Q=337$}
        \label{fig:comparison_0.03_3}
    \end{subfigure}
     \caption{Force-displacement curves for $\zeta=\SI{0.03}{mm}$ (i.e., square-like holes buckling first globally) for different numbers of basis functions $N$ for the fluctuation field. Similarly to $\zeta=\SI{-0.035}{mm}$, the ROM solution closely follows the POD solution. Both POD and ROM solutions approximate the CH2 solution poorly, implying that the training data is not representative for the global buckling of the macrostructure.} 
    \label{fig:comparison_0.03}
\end{figure}

In~\cref{fig:comparison_0.03}, a similar comparison is performed for $\zeta=\SI{0.03}{mm}$ with $N=48$, $78$ and $112$. The POD and ROM solutions match well for all $N$, but both overestimate the (post-)buckling behavior of the CH2 solution. Even though with increasing $N$ the buckling point is predicted more accurately, the postbuckling stiffness is captured poorly. This suggests that the generated training data does not properly cover the kinematics during the postbuckling stage and more representative training data is required for good approximations. Compared to the DNS solution, CH2 captures the postbuckling stiffness quite accurately, however, it overpredicts the buckling point, the prebuckling stiffness, and is again unable to detect the second buckling which occurs at around 4.5\% strain. 

\begin{figure}[th]
    \centering
    \begin{subfigure}[b]{0.25\textwidth}
        \centering
        \includegraphics[width=\textwidth]{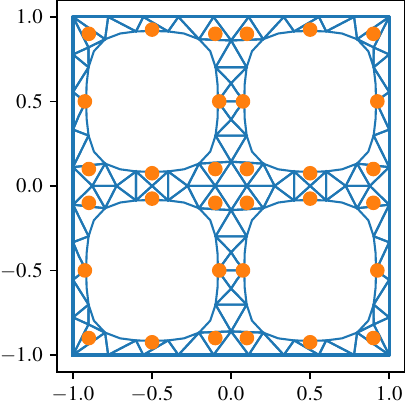}
        \caption{Coarse RVE mesh}
        \label{fig:comparison_adaptive_1}
    \end{subfigure}
    \hfill
    \begin{subfigure}[b]{0.32\textwidth}
        \centering
        \includegraphics[width=\textwidth]{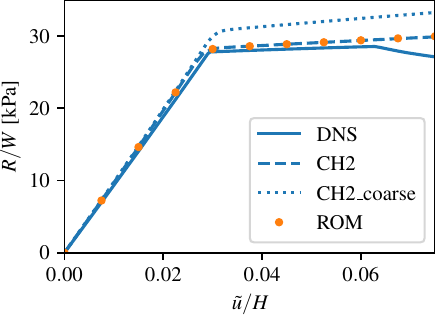}
        \caption{$\zeta=\SI{-0.035}{mm}$}
        \label{fig:comparison_adaptive_2}
    \end{subfigure}
    \hfill
    \begin{subfigure}[b]{0.32\textwidth}
        \centering
        \includegraphics[width=\textwidth]{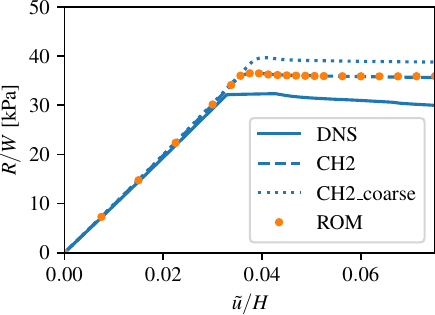}
        \caption{$\zeta=\SI{0.03}{mm}$}
        \label{fig:comparison_adaptive_3}
    \end{subfigure}
     \caption{(a) Employed coarse RVE mesh for generating a more representative training dataset. The resulting ROM has $N=48$ displacement basis functions and $Q=278$ integration points, and its solution closely follows the CH2 solution for both (b)~$\zeta=\SI{-0.035}{mm}$ and (c)~$\zeta=\SI{0.03}{mm}$. The result of the full CH2 model with the coarse RVE mesh is also shown for comparison.}
    \label{fig:comparison_adaptive}
\end{figure}

To demonstrate that the results for $\zeta=\SI{0.03}{mm}$ are improved by employing a more representative training dataset, we generated another training dataset by employing the following procedure:
\begin{enumerate}
    \item First, we solved the full CH2 problem with a coarse RVE mesh (142 six-noded elements with 746 DOFs, see~\cref{fig:comparison_adaptive_1}) for $\zeta=\{\SI{-0.05}{mm}, -\SI{0.025}{mm}, \SI{0.0}{mm}, \SI{0.025}{mm}, \SI{0.05}{mm}\}$.
    \item This way, for each value of $\zeta$ a loading trajectory of values of $\{(\Fmacro, \Gmacro)\}$ for each of the 32 macroscale quadrature points is collected.
    \item For each $\zeta$, the microscopic problem is solved with the fine mesh (see~\cref{fig:rve_parent_mesh}) along all 32 trajectories $\{(\Fmacro, \Gmacro)\}$, and snapshots of the fluctuation displacement $\wp$, weighted stress $\Wp$ and higher-order stress $\bm{\mathcal{Y}}^{\text{p}}$ are gathered.
    \item All snapshots computed with the fine mesh are utilized to construct the ROM.
\end{enumerate}
The resulting ROM for $N=48$ displacement basis functions has $Q=278$ quadrature points (with $M=28$ and $L=28$), and the resulting force-displacement curves for $\zeta=\SI{-0.035}{mm}$ and $\zeta=\SI{0.03}{mm}$ are shown in~\cref{fig:comparison_adaptive_2,fig:comparison_adaptive_3}. The ROM solution approaches the CH2 solution nearly perfectly for both cases, showing the importance of the training dataset. Additionally, the results of the full CH2 solution with the coarse RVE mesh are also shown, which shows much less accurate (post-)buckling behavior.

All simulations were executed on an Intel\textregistered~Xeon\textregistered~Platinum 8260 processor. \newtext{The run times are reported in~\cref{table:run_time_1}.} Computing the DNS solutions for $\zeta=\SI{-0.035}{mm}$ and $\SI{0.03}{mm}$ took $\SI{153}{\second}$ and $\SI{307}{\second}$ with one thread. The significant differences in computational times are caused by the global buckling, which requires many more load steps for convergence as compared to the local buckling. The ROM with $N=48$, after the offline stage is completed, took $\SI{30}{\second}$ and $\SI{69}{\second}$ for both simulations with one thread, achieving an online speed-up of 5 times as compared to the DNS solver. With $N=78$ and $N=112$, both simulations took $\SI{48}{\second}$ ($\zeta=\SI{-0.035}{mm}$) and $\SI{132}{\second}$ ($\zeta=\SI{0.03}{mm}$), and $\SI{100}{\second}$ ($\zeta=\SI{-0.035}{mm}$) and $\SI{244}{\second}$ ($\zeta=\SI{0.03}{mm}$). Concerning the offline stage of the randomly generated dataset, with one thread, 100 samples with each 20 load steps were computed in $\SI{1020}{\second}$, and constructing the ROM took another $\SI{80}{\second}$. For the more representative training dataset, the generation took significantly longer as full two-scale simulations need to be run. Note that the offline and online stage with the ROM can easily be parallelized, since all RVEs can be solved independently, which would increase the speed-up. On the other hand, DNS parallelization is less straightforward and more difficult to achieve. CH2 and POD took much longer as compared to the DNS since the considered scale separation is relatively low.

While this example problem might not be suitable for homogenization since the DNS solution can be obtained quickly, it shows that the ROM can accurately approximate the POD solution, i.e., the proposed algorithm for finding a sparse integration scheme works well. Moreover, the POD solution approaches the CH2 solution (provided the training data is representative), which in turn approximates the DNS well.

\begin{table}[th]
\centering
\caption{\newtext{Run times for DNS and three ROM solutions with $N=48$, $N=78$ and $N=112$. As this problem is quite small, the DNS can be computed efficiently, and the speed ups are not that high for the ROMs.}}
\newtext{\scalebox{1.2}{\begin{tabular}{c|cccc}
           & DNS                   & ROM48               & ROM78               & ROM112              \\ \hline
$\zeta=\SI{-0.035}{mm}$ & $\SI{153}{\second}$ & $\SI{30}{\second}$ & $\SI{48}{\second}$ & $\SI{100}{\second}$ \\
$\zeta=\SI{0.03}{mm}$ & $\SI{307}{\second}$ & $\SI{69}{\second}$ & $\SI{132}{\second}$ & $\SI{244}{\second}$
\end{tabular}}}
\label{table:run_time_1}
\end{table}

\subsection{Biaxial Compression of Graded Cruciform}\label{subsec:ex2}
\begin{figure}[th]
    \centering
    \begin{subfigure}[b]{0.415\textwidth}
        \centering
        \includegraphics[width=\textwidth]{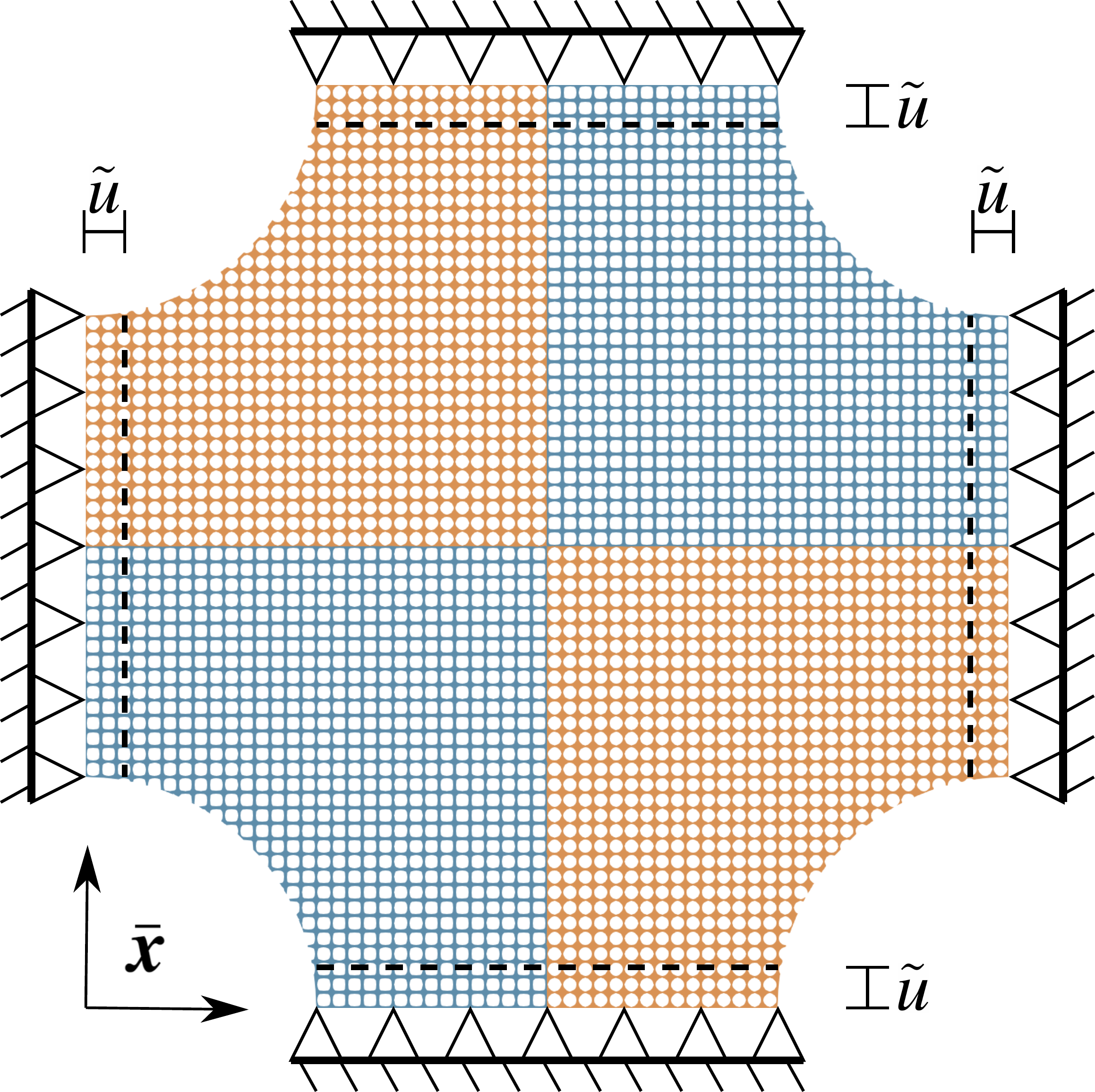}
        \caption{Geometry 1}
        \label{fig:cruciform_dns1}
    \end{subfigure}
    \qquad
    \begin{subfigure}[b]{0.415\textwidth}
        \centering
        \includegraphics[width=\textwidth]{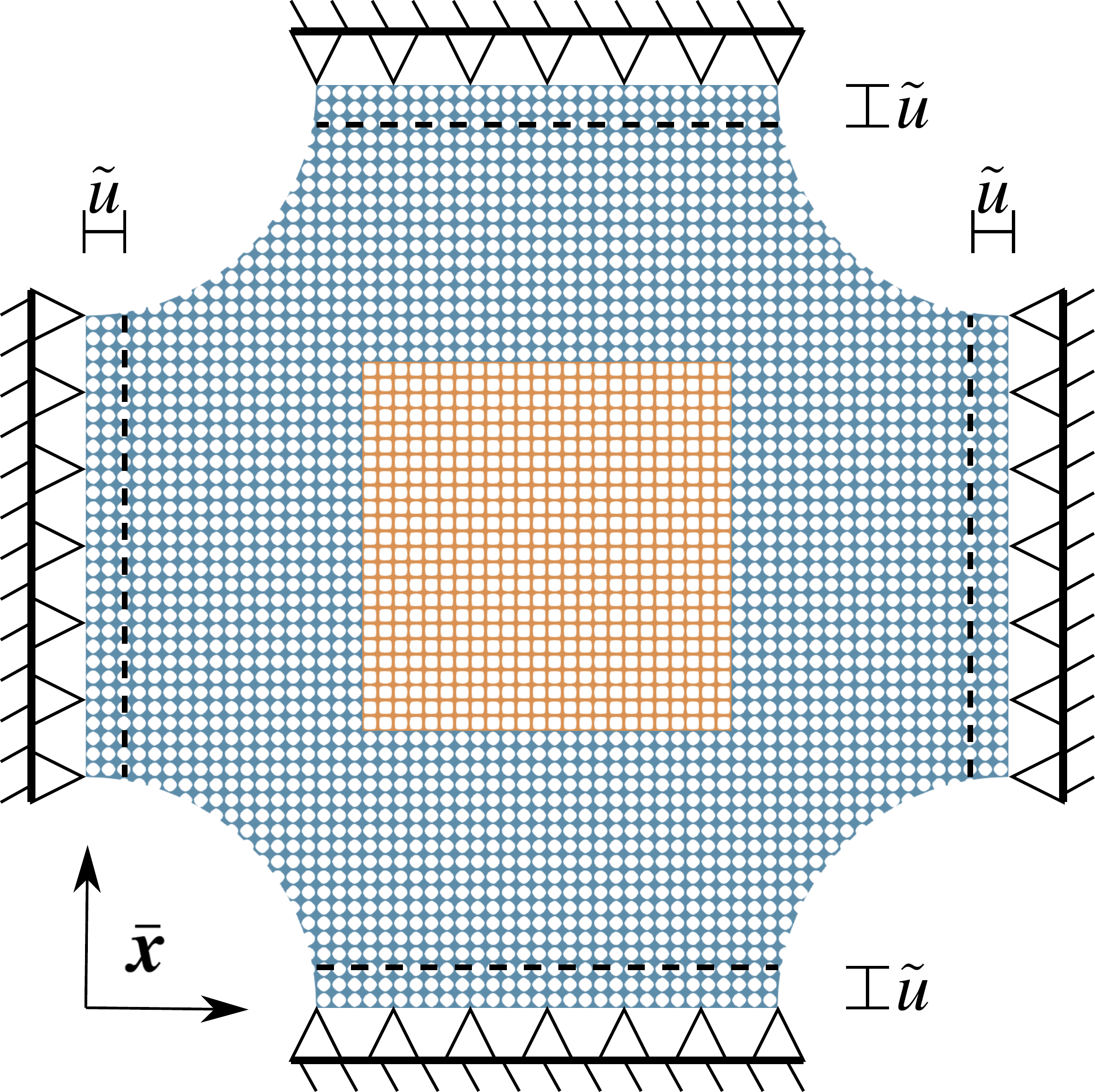}
        \caption{Geometry 2}
        \label{fig:cruciform_dns2}
    \end{subfigure}
     \caption{Two geometries are solved with the DNS solver. (a) $\zeta=\SI{0.03}{mm}$ is set for the top right and bottom left part (in blue), and $\zeta=\SI{-0.05}{mm}$ for the top left and bottom right part (orange). (b) $\zeta=\SI{0.05}{mm}$ is set in the center (orange) for $\Xmacro\in[\SI{18}{mm},\SI{42}{mm}]\times[\SI{18}{mm},\SI{42}{mm}]$ and $\zeta=\SI{-0.075}{mm}$ elsewhere (blue). Six-noded triangular elements are employed for both geometries, resulting in (a) 3,475,044 DOFs and 800,889 elements, and (b) 3,627,610 DOFs and 839,580 elements.}
    \label{fig:cruciform_dns}
\end{figure}

\begin{figure}[th]
    \centering
    \begin{subfigure}[b]{0.35\textwidth}
        \centering
        \includegraphics[width=\textwidth]{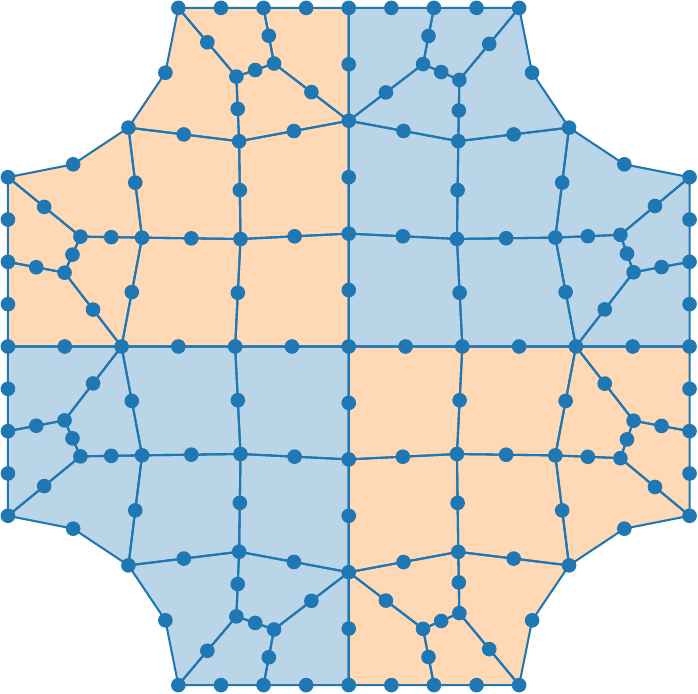}
        \caption{Geometry 1}
        \label{fig:cruciform_rom1}
    \end{subfigure}
    \qquad
    \begin{subfigure}[b]{0.35\textwidth}
        \centering
        \includegraphics[width=\textwidth]{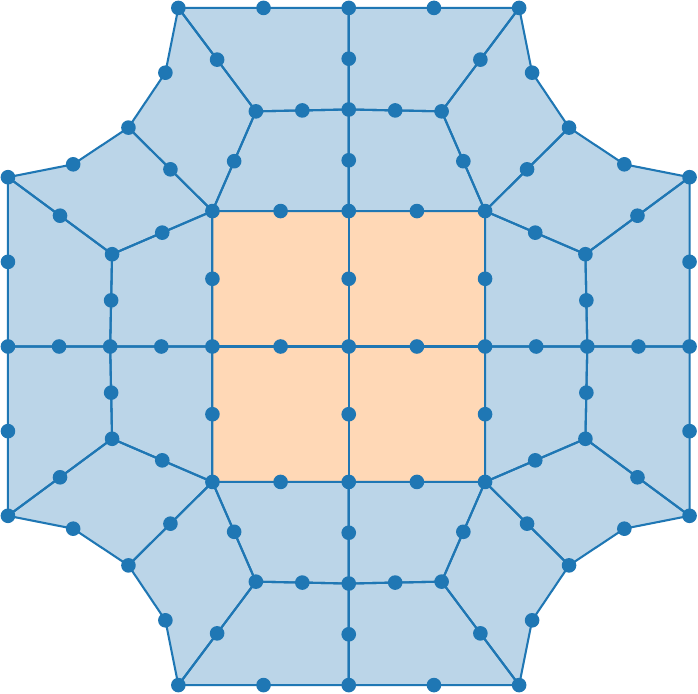}
        \caption{Geometry 2}
        \label{fig:cruciform_rom2}
    \end{subfigure}
     \caption{Discretization of the homogenized cruciform used for CH2 and ROM. (a) $\zeta=\SI{0.03}{mm}$ is set for the blue elements and $\zeta=\SI{-0.05}{mm}$ for the orange elements. (b) The orange and blue elements correspond to $\zeta=\SI{0.05}{mm}$ and $\zeta=\SI{-0.075}{mm}$, respectively. The meshes have (a) 48 and (b) 28 elements with four quadrature points each.}
    \label{fig:cruciform_rom}
\end{figure}

The second example deals with the biaxial compression of a graded cruciform-shaped macrostructure with varying hole shapes (i.e., spatially varying $\zeta$ field) throughout the domain, see~\cref{fig:cruciform_dns}. Each side edge has length $\SI{30}{mm}$, and the cut out parts at each corner are quarter circles with a radius of $\SI{15}{mm}$. Both example parameterizations shown in~\cref{fig:cruciform_dns} are considered and computed with the DNS, CH2 and the ROM solver. The discretized DNS problem has for~\cref{fig:cruciform_dns1} 3,475,044 DOFs and 800,889 elements, and 3,627,610 DOFs and 839,580 elements for~\cref{fig:cruciform_dns2}. Each side edge is compressed by 2\% in the normal direction, while being fixed in the tangential direction. For CH2 and the ROM, additionally, the $xx$- and $yx$-components of the deformation gradient are fixed to $\widehat{\macro{F}}_{xx}=1$ and $\widehat{\macro{F}}_{yx}=0$ on the top and bottom horizontal edges, and the $xy$- and $yy$-components of the deformation gradient are fixed to $\widehat{\macro{F}}_{xy}=0$ and $\widehat{\macro{F}}_{yy}=1$ on the left and right vertical edges. The simulation meshes employed for CH2 and the ROM are shown in~\cref{fig:cruciform_rom}.

Since the deformation of this example is similar to the previous example in~\cref{subsec:ex1}, the already trained ROM (with randomly generated training data, cf.~\cref{table:rve_bounds1}) is re-used here with $N=48$, $78$ and $112$ displacement basis functions with $Q=297$, $318$ and $337$ integration points, and referred to as ROM48, ROM78 and ROM112.

\subsubsection{Results}
\begin{figure}[th]
    \centering
    \begin{subfigure}[b]{0.45\textwidth}
        \centering
        \includegraphics[width=\textwidth]{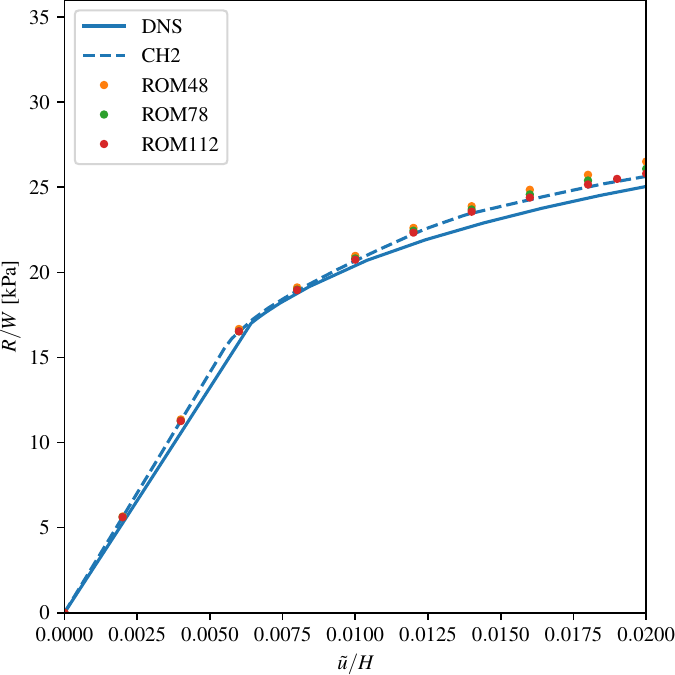}
        \caption{Geometry 1}
        \label{fig:cruciform_comparison1}
    \end{subfigure}
    \qquad
    \begin{subfigure}[b]{0.45\textwidth}
        \centering
        \includegraphics[width=\textwidth]{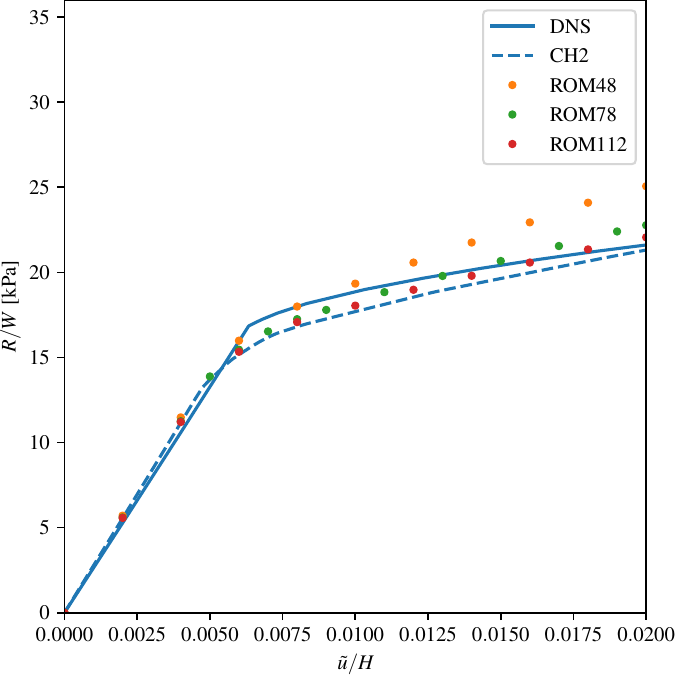}
        \caption{Geometry 2}
        \label{fig:cruciform_comparison2}
    \end{subfigure}
     \caption{Force-displacement curves for both example geometries of~\cref{fig:cruciform_dns,fig:cruciform_rom} obtained for DNS, CH2, and three ROMs with different numbers of basis functions $N$ and quadrature points $Q$. (a)~All ROM solutions are close to the CH2 solution. The CH2 solution approximates the DNS adequately. (b) The ROMs do not recover the CH2 solution accurately, since $\zeta=\SI{-0.075}{mm}$ is outside the training data. For higher numbers of basis functions, the approximation gets increasingly more accurate, and for ROM112 only small deviations are observed during the postbuckling stage. CH2 is unable to predict the correct buckling load.}
    \label{fig:cruciform_comparison}
\end{figure}

\begin{figure}[th]
    \centering
    \begin{subfigure}[b]{0.3357\textwidth}
        \centering
        \includegraphics[width=\textwidth]{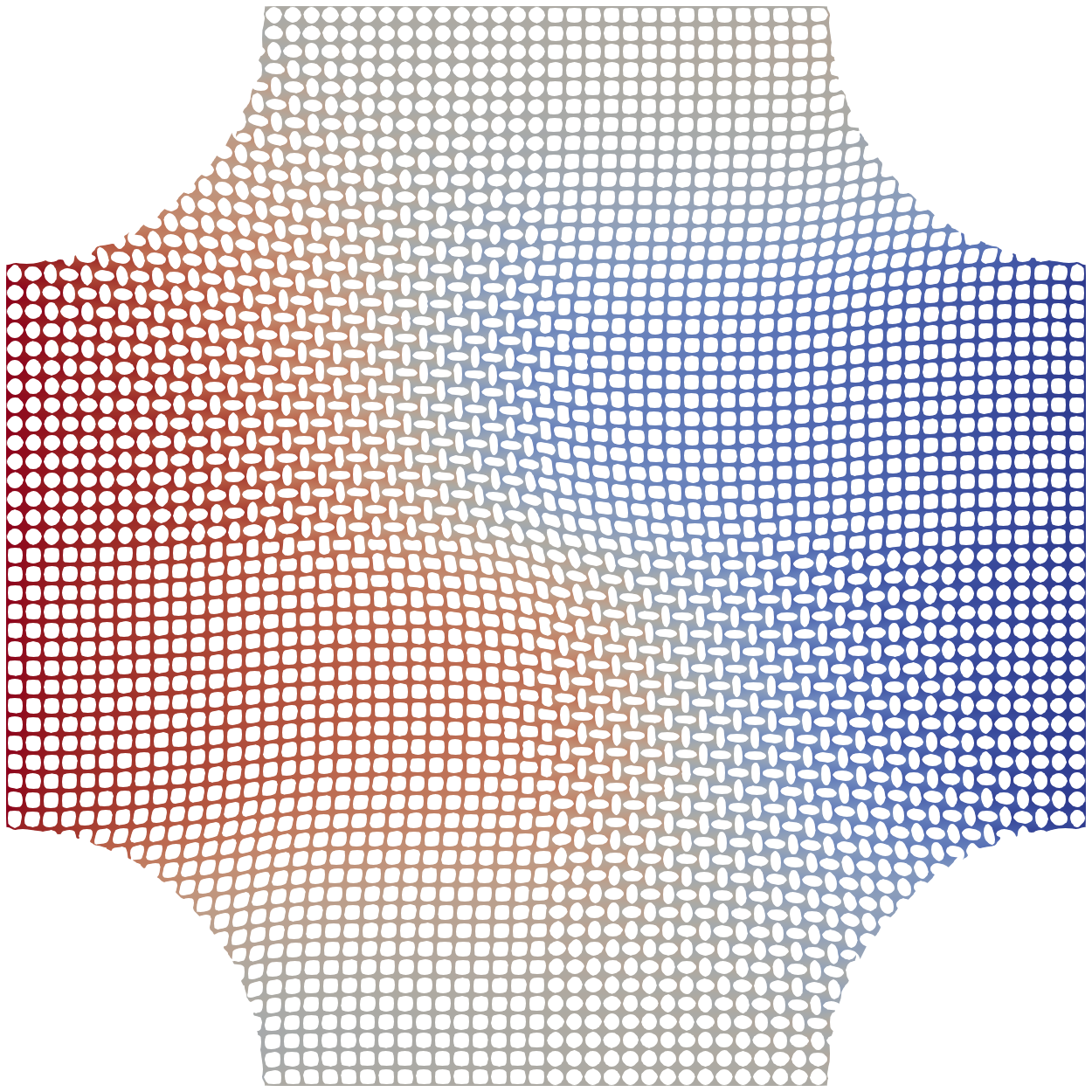}
        \label{fig:cruciform1_kinematics1}
    \end{subfigure}
    \qquad
    \begin{subfigure}[b]{0.3357\textwidth}
        \centering
        \includegraphics[width=\textwidth]{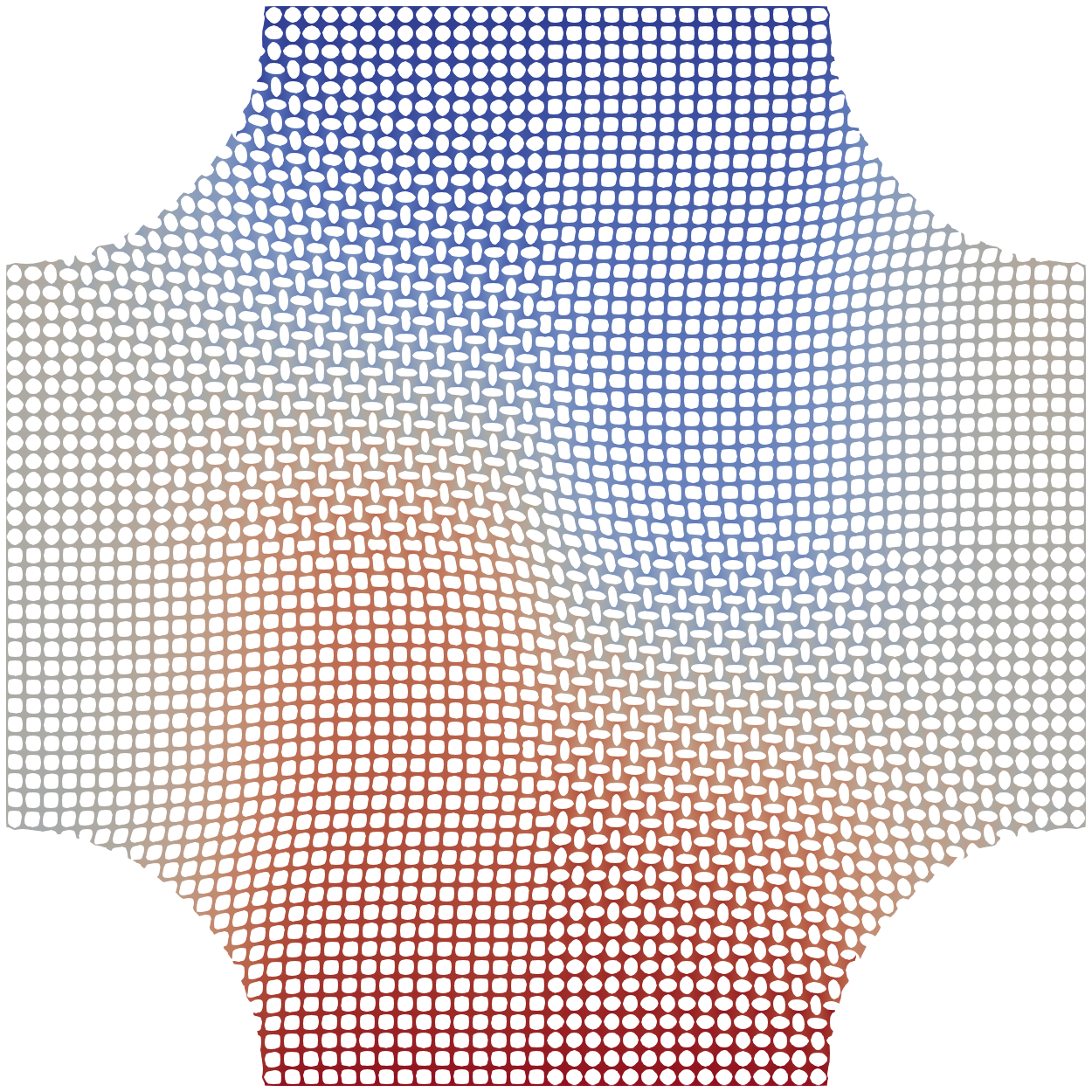}
        \label{fig:cruciform1_kinematics2}
    \end{subfigure}
    \begin{subfigure}[b]{0.47\textwidth}
        \centering
        \includegraphics[width=\textwidth]{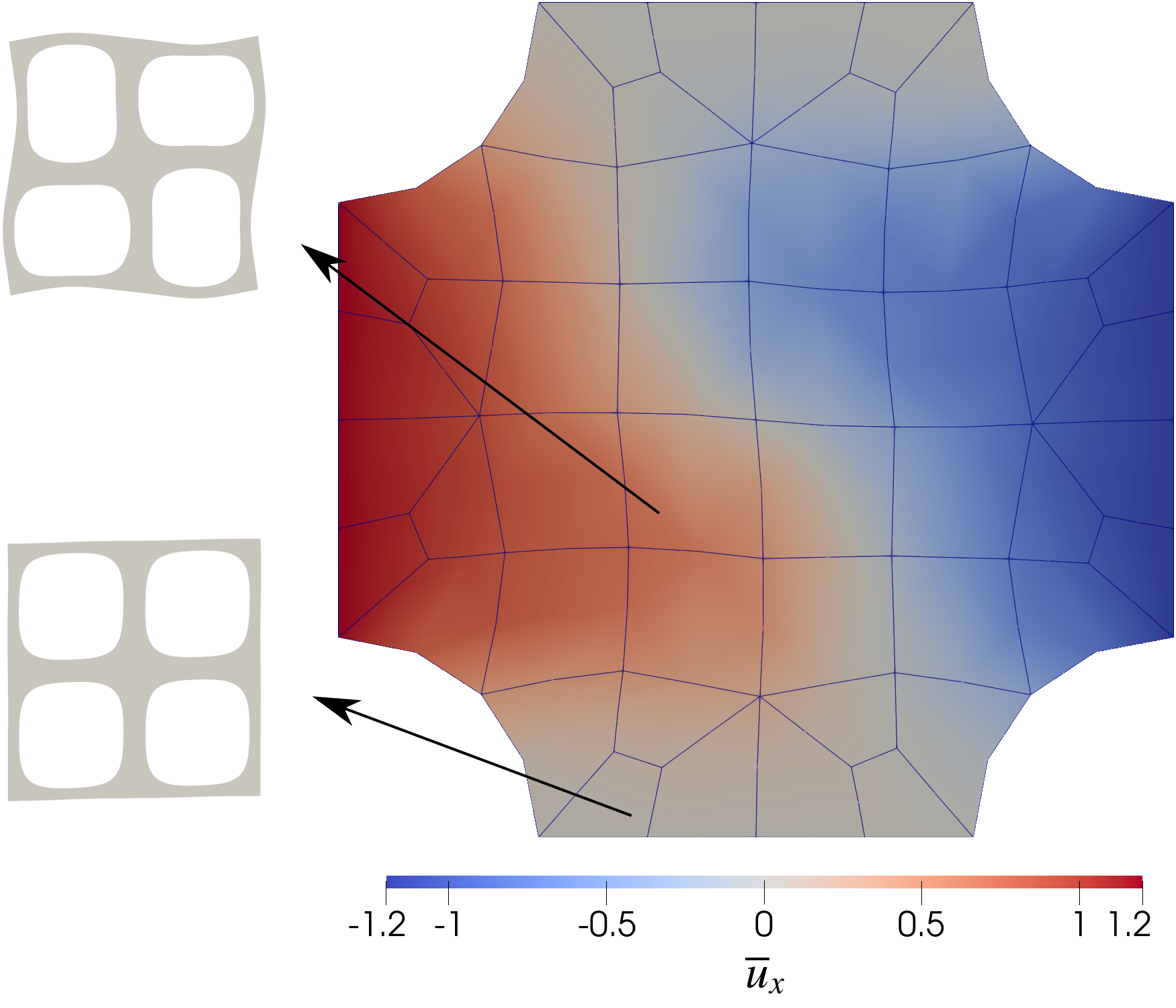}
        \label{fig:cruciform1_kinematics3}
    \end{subfigure}
    \qquad
    \begin{subfigure}[b]{0.47\textwidth}
        \centering
        \includegraphics[width=\textwidth]{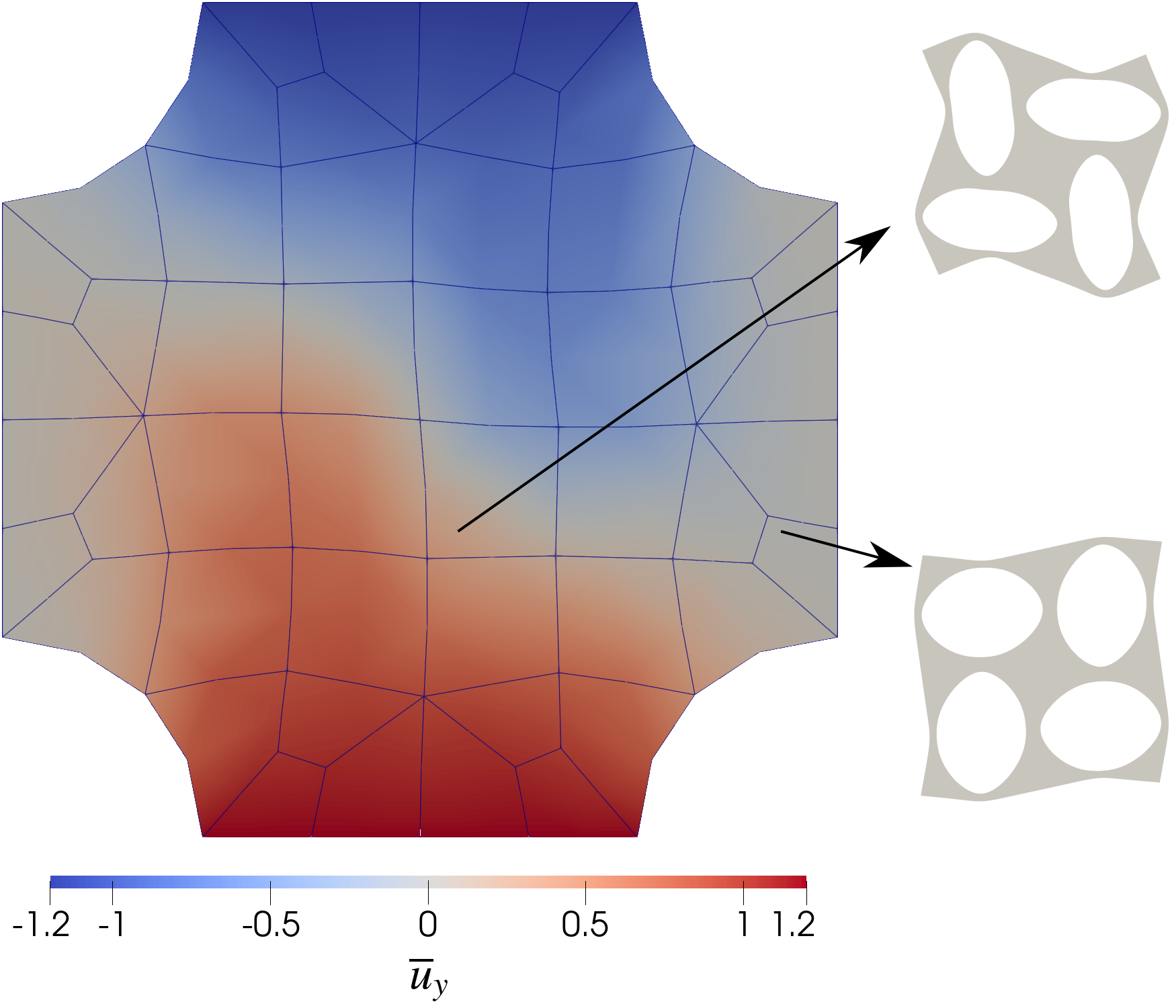}
        \label{fig:cruciform1_kinematics4}
    \end{subfigure}
     \caption{Displacement fields for geometry 1 obtained with DNS (top row) and with ROM48 (bottom row). The shear band forming along the diagonal is clearly captured.}
    \label{fig:cruciform1_kinematics}
\end{figure}

\begin{figure}[th]
    \centering
    \begin{subfigure}[b]{0.3357\textwidth}
        \centering
        \includegraphics[width=\textwidth]{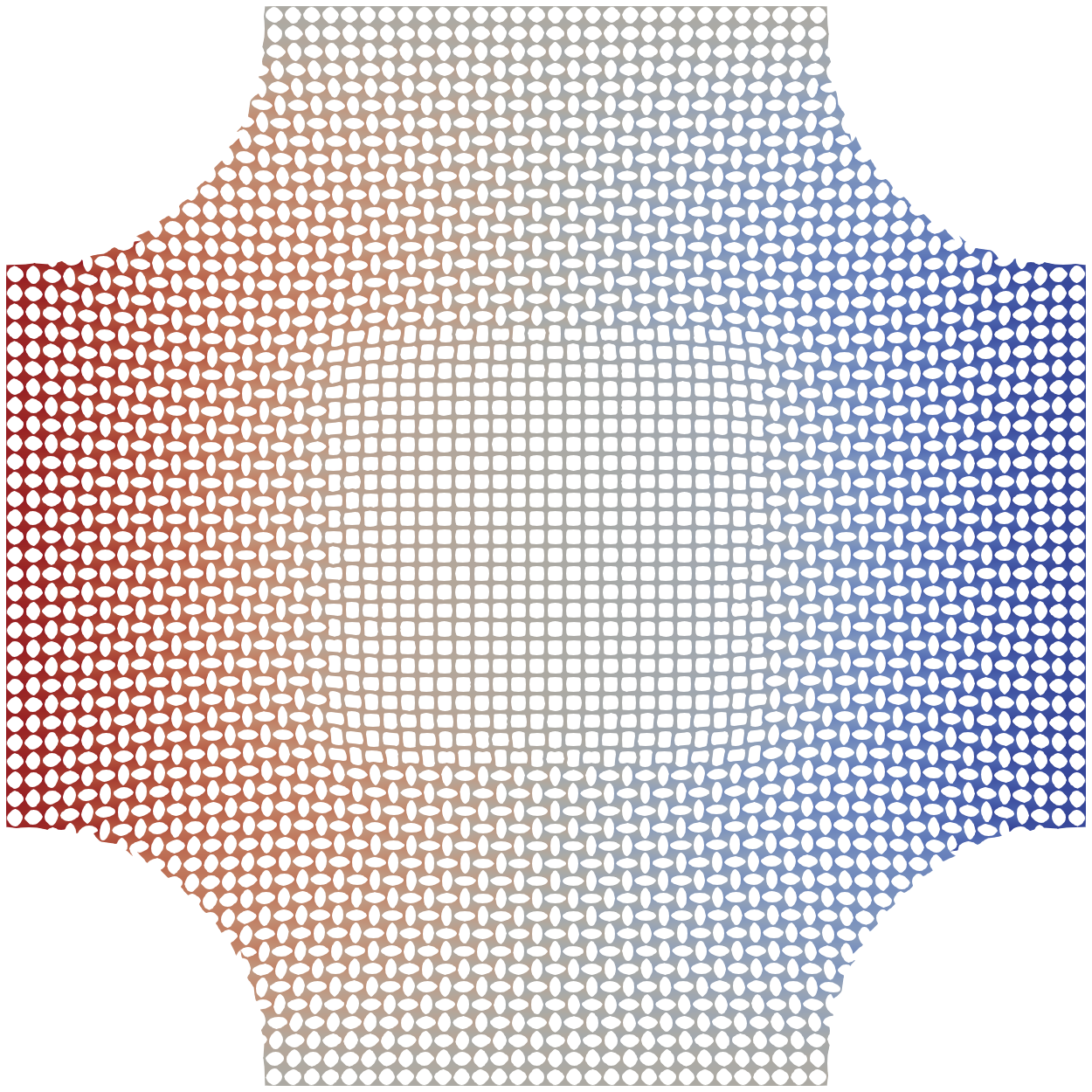}
        \label{fig:cruciform2_kinematics1}
    \end{subfigure}
    \qquad
    \begin{subfigure}[b]{0.3357\textwidth}
        \centering
        \includegraphics[width=\textwidth]{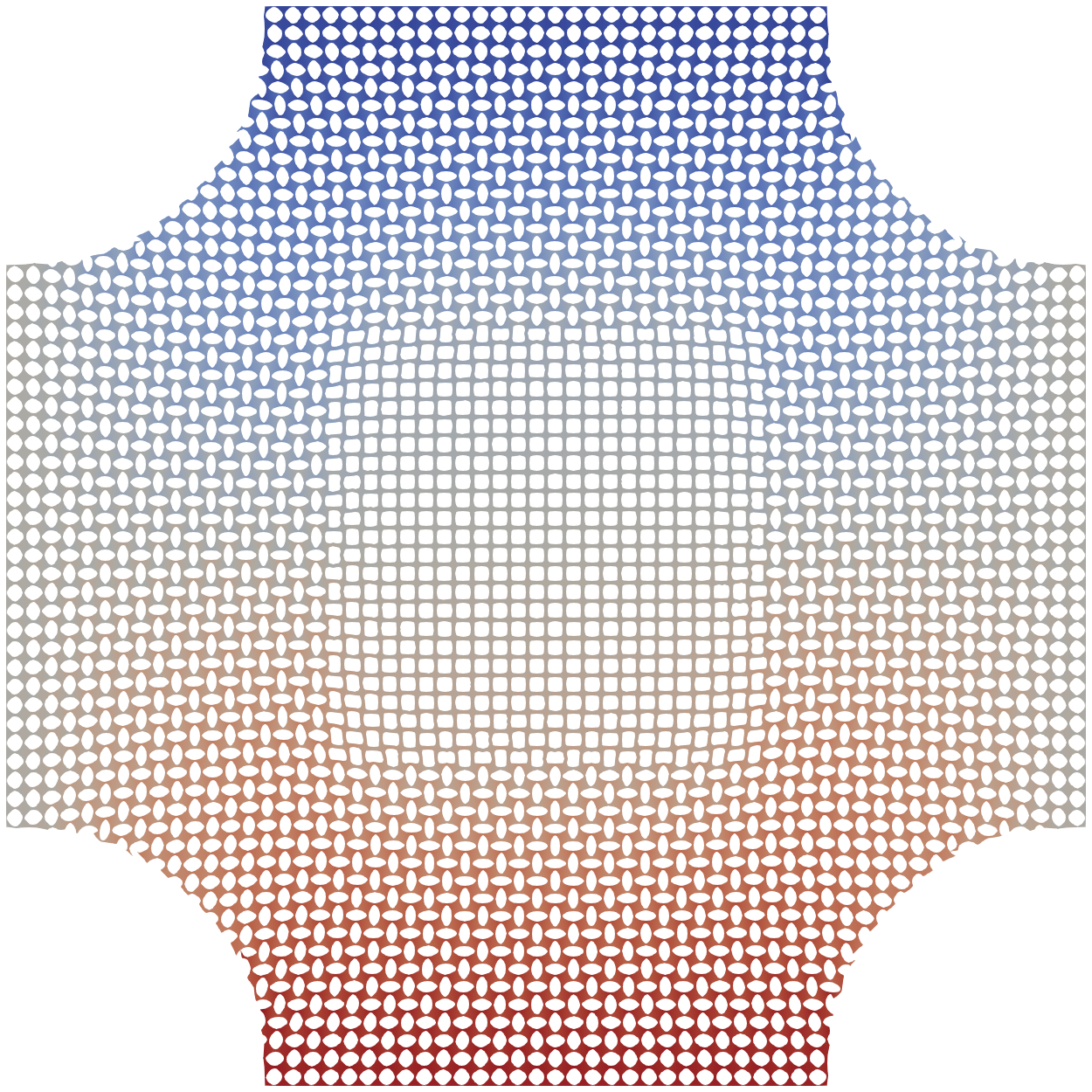}
        \label{fig:cruciform2_kinematics2}
    \end{subfigure}
    \begin{subfigure}[b]{0.47\textwidth}
        \centering
        \includegraphics[width=\textwidth]{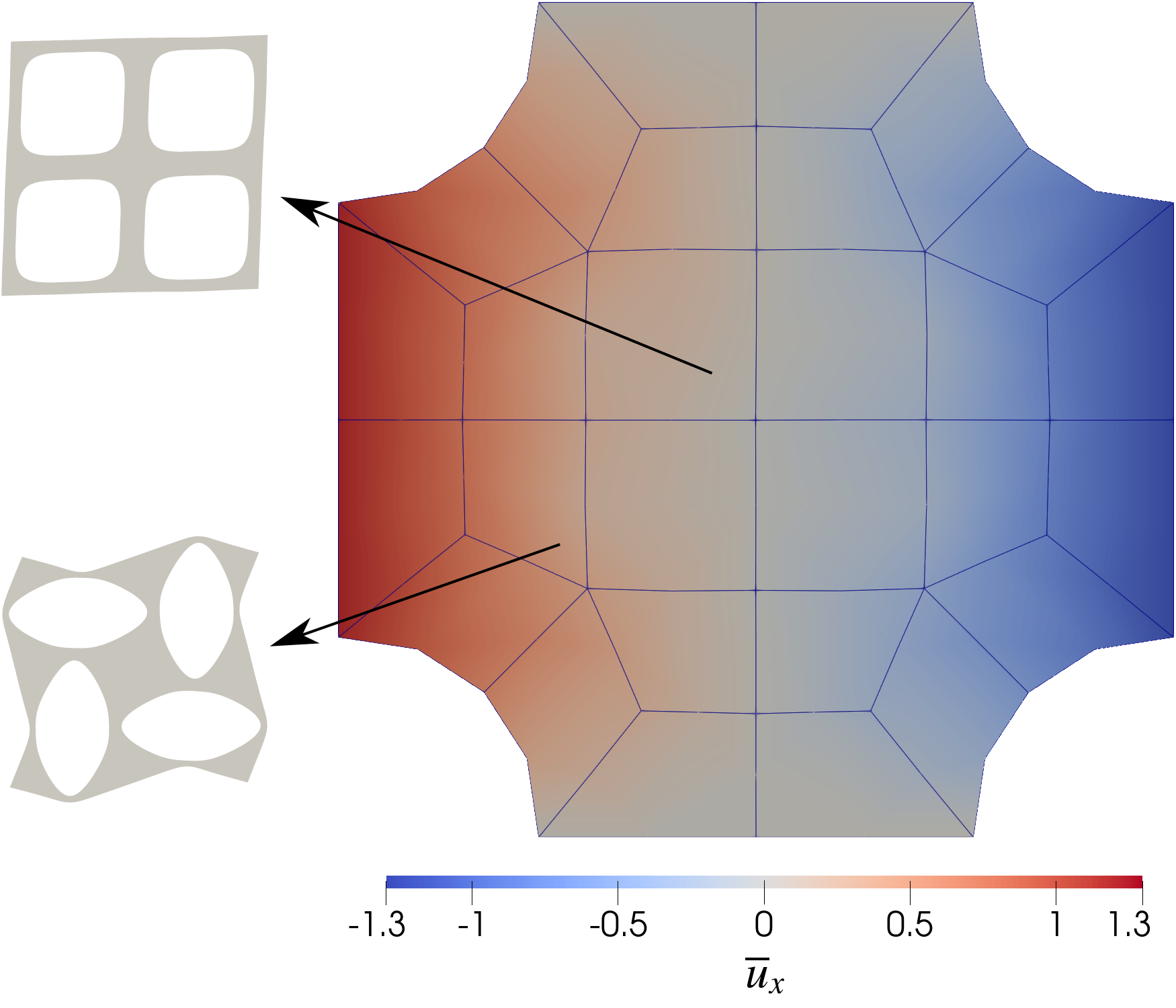}
        \label{fig:cruciform2_kinematics3}
    \end{subfigure}
    \qquad
    \begin{subfigure}[b]{0.47\textwidth}
        \centering
        \includegraphics[width=\textwidth]{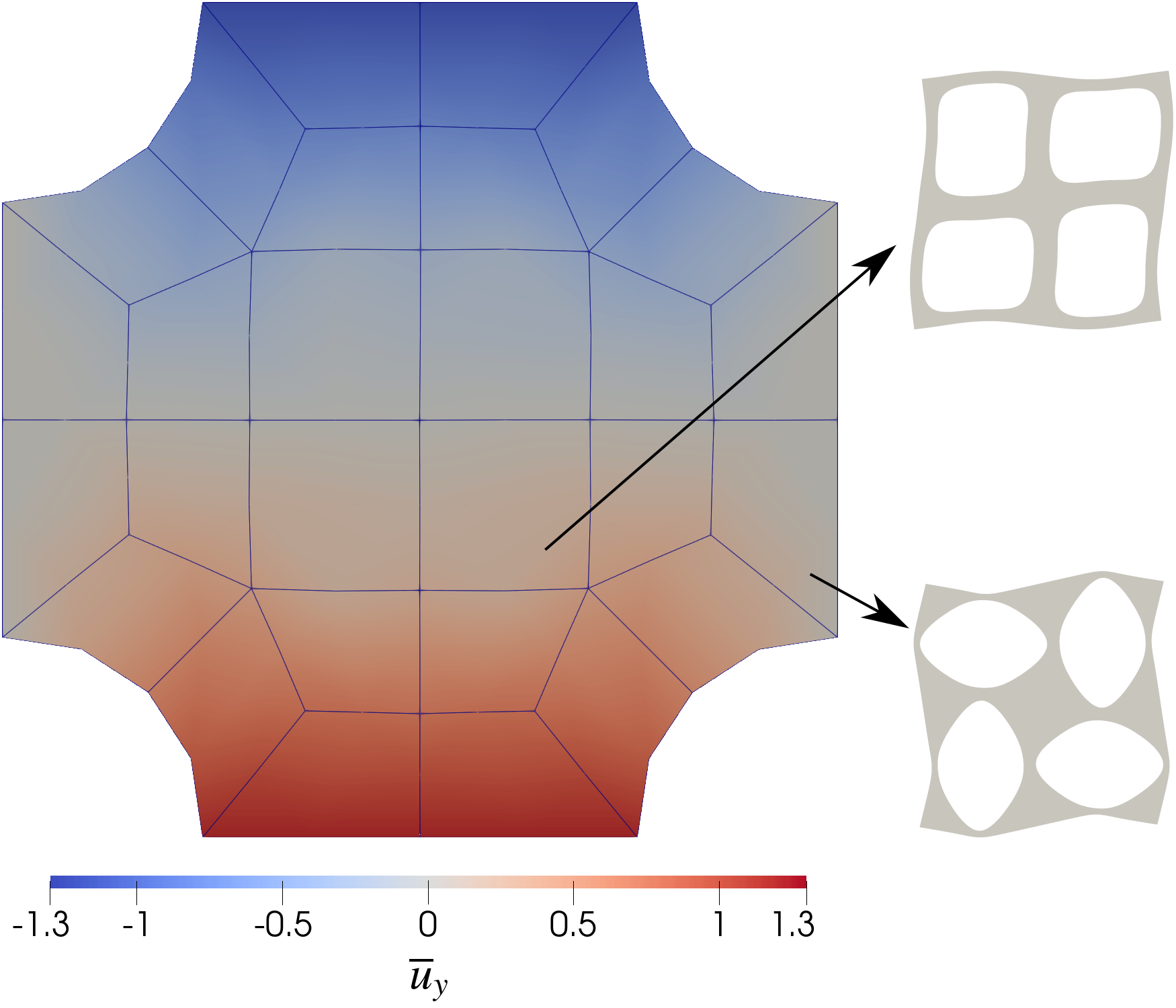}
        \label{fig:cruciform2_kinematics4}
    \end{subfigure}
     \caption{Displacement fields for geometry 2 obtained with DNS (top row) and with ROM48 (bottom row). The sharp transition in the center due to the sharp change in $\zeta$ is more or less captured.}
    \label{fig:cruciform2_kinematics}
\end{figure}

Both example geometries are solved with DNS, CH2 and the three ROM solvers (ROM48, ROM78 and ROM112). To compare the results, the total resulting reaction force $R$ acting on the top edge, normalized with the length of the edge $W=\SI{30}{mm}$, is plotted over the applied displacement $\tilde{u}$, normalized with the total height $H=\SI{60}{mm}$, in~\cref{fig:cruciform_comparison}. For the first geometry (see~\cref{fig:cruciform_comparison1}), all ROM solutions recover the CH2 solution nearly perfectly. As compared to the DNS solution, CH2 again predicts a higher prebuckling stiffness, but the buckling load and postbuckling stage are predicted quite well. For the second geometry (see~\cref{fig:cruciform_comparison2}), CH2 cannot capture the correct buckling load. Since in this example, the center part is set to $\zeta=\SI{-0.075}{mm}$, which is outside the training data (sampled from $\zeta=\{\SI{-0.05}{mm}, -\SI{0.025}{mm}, \SI{0.0}{mm}, \SI{0.025}{mm}, \SI{0.05}{mm}\}$), ROM48 is not able to follow the CH2 solutions closely. The error reduces with $N=78$ and $112$. For both geometries, both components of the displacement field at the final loading are shown for the DNS and ROM48 solutions in~\cref{fig:cruciform1_kinematics,fig:cruciform2_kinematics}, where clearly the trend and magnitudes of the displacement fields are comparable. Deformed RVEs at selected quadrature points of the ROM48 solutions are also shown, which behave similarly to the DNS at similar location.

\newtext{The run times (using one thread of Intel\textregistered~Xeon\textregistered~Platinum 8260) are reported in~\cref{table:run_time_2}.} Solving both problems with the DNS solver took $\SI{15129}{\second}$ and $\SI{14087}{\second}$, while with ROM48 it took $\SI{164}{\second}$ and $\SI{101}{\second}$, implying online speed-ups of $92$ 
and $139$ times. With ROM78, the run times were $\SI{369}{\second}$ and $\SI{333}{\second}$, meaning speed-ups of $41$ and $42$ times. With ROM112, the solution took $\SI{975}{\second}$ and $\SI{439}{\second}$, which still amounts to online speed-ups of $15$ and $32$ times. The computational costs of the offline stage are the same as reported for the previous example in~\cref{subsec:ex1}. The obtained speed-ups could be greatly increased by using more threads due to the superior scaling of the multiscale formulation over the DNS. The run times of CH2 are again much higher than the run times of DNS.

\begin{table}[th]
\centering
\caption{\newtext{Run times for DNS and three ROM solutions with $N=48$, $N=78$ and $N=112$. For all ROMs, large online speed ups are observed for both geometries.}}
\newtext{\scalebox{1.2}{\begin{tabular}{c|cccc}
           & DNS                   & ROM48               & ROM78               & ROM112              \\ \hline
Geometry 1 & $\SI{15129}{\second}$ & $\SI{164}{\second}$ & $\SI{369}{\second}$ & $\SI{975}{\second}$ \\
Geometry 2 & $\SI{14087}{\second}$ & $\SI{101}{\second}$ & $\SI{333}{\second}$ & $\SI{439}{\second}$
\end{tabular}}}
\label{table:run_time_2}
\end{table}

The DNS solution took several hours (with one thread), mostly because of the detection of instabilities (i.e., checking the system matrix for negative eigenvalues and eigenvalues close to zero). If a large-scale problem (in 3D) was considered, the DNS solution might become infeasible, since (1) detecting negative eigenvalues is computationally expensive, and (2) negative eigenvalues of the system matrix may cause problems for iterative solvers, while direct solvers become too computationally expensive for such large systems. On the other hand, the ROM solution should remain relatively computationally inexpensive, since the solver can be easily parallelized by solving all RVE problems at the macroscopic integration points in parallel. An additional advantage of the ROM is that, after training, different geometrical parameters inside the macrostructure can be easily tuned, while for the DNS, the meshing can become expensive and challenging, especially for 3D problems. This makes this ROM an interesting candidate for the material design of buckling structures.

\section{Conclusions}\label{sec5}
In this work, we proposed a reduced-order model (ROM) for second-order computational homogenization (CH2), based on proper orthogonal decomposition and a novel hyperreduction method that uses ideas from the empirical cubature method that is specifically suited for CH2. Several aspects on the derivation of the reduced system, including the treatment of constraints and geometrical parameterizations, expressions for the effective quantities, and the novel hyperreduction algorithm were discussed. Afterwards, the ROM was tested on two numerical examples, in which the macrostructures are compressed and multiscale buckling occurs. The ROM solutions were critically evaluated by comparison against the results obtained by direct numerical simulation (DNS) and the full CH2 model. The first example demonstrated that the proposed hyperreduction algorithm discovered integration points and weights that yield accurate results for several parameterizations of the microstructure. The second example concerned a more complex application, in which the geometry of the microstructures is varied within the macroscopic domain, and for which the DNS solution takes a substantial amount of time to compute. When employing the ROM for this problem, speed-ups ranging from 15 to 139 as compared to the DNS were achieved with one thread. These speed-ups could be further increased by employing more threads, since, in general, the multiscale problem scales much better than the DNS.

To the best of our knowledge, this work is the first attempt of accelerating an enriched computational homogenization formulation. Although we proposed a reduced-order model for CH2, we are confident that our findings and employed methods also extend to other formulations, e.g., based on micromorphic computational homogenization. \newtext{It would also be interesting to test the methodology on more complicated material models, such as damage and fracture.} As different parameterizations of the microstructure can be treated as well, interesting applications can be realized with this framework, such as two-scale shape optimization problems, design of materials and uncertainty quantification.

\section*{Acknowledgments}
This result is part of a project that has received funding from the European Research Council (ERC) under the European Union’s Horizon 2020 Research and Innovation Programme (Grant Agreement No. 818473).

\section*{Data Availability}
The data that support the findings of this study are available from the corresponding author upon request.

\section*{Conflict of interest}
The authors declare no potential conflict of interests.

\bibliography{references}


\end{document}